\documentclass[12pt,a4paper]{article}
\usepackage{amsmath,amssymb,graphicx}

\makeatletter
\newif\if@preliminary
\@preliminaryfalse
\def\preliminary{\@preliminarytrue}
\def\preprintno#1{\def\@preprintno{#1}}
\def\address#1{\def\@address{#1}}
\def\email#1#2{\thanks{\tt #1@{}#2}}
\def\abstract#1{\def\@abstract{#1}}
\renewcommand\abstractname{ABSTRACT}
\newlength\preprintnoskip
\newlength\abstractwidth
\@titlepagetrue
\renewcommand\maketitle{\begin{titlepage}%
  \let\footnotesize\small
  \hfill\parbox{\preprintnoskip}{%
  \begin{flushright}\@preprintno\end{flushright}}\hspace*{1cm}
  \vskip 60\p@
  \begin{center}%
    {\Large\bf\boldmath \@title \par}\vskip 1cm%
    {\sc\@author \par}\vskip 3mm%
    {\@address \par}%
    \if@preliminary
      \vskip 2cm {\large\sf PRELIMINARY DRAFT \par \@date}%
    \fi
  \end{center}\par
  \@thanks
  \vfill
  \begin{center}%
    \parbox{\abstractwidth}{\centerline{\abstractname}%
    \vskip 3mm%
    \@abstract}
  \end{center}
  \end{titlepage}%
  \setcounter{footnote}{0}%
  \let\thanks\relax\let\maketitle\relax
  \gdef\@thanks{}\gdef\@author{}\gdef\@address{}%
  \gdef\@title{}\gdef\@abstract{}\gdef\@preprintno{}
}%
\long\def\@makecaption#1#2{%
  \vskip\abovecaptionskip
  \sbox\@tempboxa{#1: \emph{#2}}%
  \ifdim \wd\@tempboxa >\hsize
    #1: \emph{#2}\par
  \else
    \hbox to\hsize{\hfil\box\@tempboxa\hfil}%
  \fi
  \vskip\belowcaptionskip}
\def\fmslash{\@ifnextchar[{\fmsl@sh}{\fmsl@sh[0mu]}}
\def\fmsl@sh[#1]#2{%
  \mathchoice
    {\@fmsl@sh\displaystyle{#1}{#2}}%
    {\@fmsl@sh\textstyle{#1}{#2}}%
    {\@fmsl@sh\scriptstyle{#1}{#2}}%
    {\@fmsl@sh\scriptscriptstyle{#1}{#2}}}
\def\@fmsl@sh#1#2#3{\m@th\ooalign{$\hfil#1\mkern#2/\hfil$\crcr$#1#3$}}
\makeatother

\def\ie{{\it i.e. }}
\def\eg{{\it e.g. }}

\newcommand{\sms}{\widetilde m}
\newcommand{\mGUT}{M_{\rm GUT}}

\newcommand{\MeV}{{\ensuremath\rm \, MeV}}
\newcommand{\GeV}{{\ensuremath\rm \, GeV}}
\newcommand{\TeV}{{\ensuremath\rm \, TeV}}

\newcommand{\s}{{\ensuremath\rm s}}
\newcommand{\fb}{{\ensuremath\rm \, fb}}
\newcommand{\ab}{{\ensuremath\rm \, ab}}
\newcommand{\pb}{{\ensuremath\rm \, pb}}

\newcommand{\nne}{\tilde{\chi}_1^0}
\newcommand{\nnz}{\tilde{\chi}_2^0}
\newcommand{\nnd}{\tilde{\chi}_3^0}
\newcommand{\nnv}{\tilde{\chi}_4^0}

\newcommand{\cpe}{\tilde{\chi}_1^+}
\newcommand{\cpz}{\tilde{\chi}_2^+}

\newcommand{\cme}{\tilde{\chi}_1^-}
\newcommand{\cmz}{\tilde{\chi}_2^-}
\begin{document}
\preprintno{DESY 04-136\\CERN-PH-TH/2004-144\\IPPP/04/46\\DCPT/04/92\\hep-ph/0408088\\[0.5\baselineskip] August 2004}
\title{%
 Split Supersymmetry at Colliders
}
\author{%
 W.~Kilian\email{wolfgang.kilian}{desy.de}${}^a$,
 T.~Plehn\email{tilman.plehn}{cern.ch}${}^b$,
 P.~Richardson\email{Peter.Richardson}{durham.ac.uk}${}^c$
 and
 E.~Schmidt\email{schmidt}{rubin.physik2.uni-rostock.de}${}^d$
}
\address{\it%
 ${}^a$Deutsches Elektronen-Synchrotron DESY, D--22603 Hamburg, Germany\\
 ${}^b$CERN, CH--1211 Geneva 23, Switzerland\\
 ${}^c$Institute for Particle Physics Phenomenology, University of
  Durham, DH1 3LE, UK\\
 ${}^d$Fachbereich Physik, University of Rostock, D-18051 Rostock, Germany
}

\abstract{We consider the collider phenomenology of
  split-supersymmetry models. Despite the challenging
  nature of the signals in these models the long-lived gluino can be
  discovered with masses above $2\,\TeV$ at the LHC. At a
  future linear collider we will be able to observe the
  renormalization group effects from split supersymmetry, using
  measurements of the neutralino and chargino masses and cross
  sections.}

\maketitle


\section{Introduction}

The standard signatures of supersymmetry at a hadron collider consist
of multi jet and multi lepton final states with missing transverse
energy~\cite{atlastdrII}.
The underlying physics typically involves
pair production of new heavy coloured particles (squarks and gluinos),
which cascade-decay into the lightest supersymmetric particle~(LSP).
This particle, also a dark matter
candidate, leaves the detector undetected.  In standard supergravity-mediated
models~\cite{Alvarez-Gaume:1983gj}, it is the lightest neutralino,
while in gauge-mediated supersymmetry-breaking
models~\cite{Dine:1981za} the gravitino plays this role.

This type of phenomenology naturally appears in hidden-sector models
of supersymmetry breaking.  If there is a large mass gap between the
electroweak scale and the lowest new-physics scale, where new particles
interact directly with the visible sector,
the MSSM is the correct effective theory over many orders of
magnitude (modulo fine-tuning arguments).
In the MSSM, the renormalization group flow drives the
masses of the coloured particles to comparatively large values,
while the weakly interacting
states stay relatively light, but all particles are expected to have masses
below a few TeV~\cite{Drees:1995hj}.\smallskip

However, the actual supersymmetric spectrum need not follow this
generic expectation.  For instance, there are scenarios where the
gluino is the LSP~\cite{Farrar:1978xj,Baer:1998pg}.  If the gluino is long-lived, it
will pick up quarks and gluons from the vacuum and hadronize into a
(meta)stable $R$-hadron~\cite{Farrar:1978xj}.  In that case,
the SUSY signal no longer consists of missing energy in the hard
process.  Instead, there will be atypical hits in the hadronic
calorimeter (and in other parts of the detector). These correspond to
a $R$-hadron, which is either stopped in or passes through the detector,
possibly leaving a fake missing-energy signal.

Recently, this feature has appeared in the context of \emph{split
supersymmetry} (SpS) models~\cite{Arkani-Hamed:2004fb,Giudice:2004tc}.
It is a well-known fact that all known models of electroweak symmetry
breaking, including supersymmetric ones, require an incredible amount
of fine-tuning of the vacuum energy, such that the resulting
cosmological constant is as small as observed.
Weakly interacting models, which contain a Higgs boson,
require somewhat less, but still incredible additional fine-tuning of the
electroweak scale.  The latter hierarchy problem is ameliorated in
models with exact cancellations in the Higgs sector quantum
corrections due to TeV-scale new particles.  Softly broken
supersymmetry achieves this to all orders in the coupling constants.
However, if we accept the fine-tuning of the vacuum energy without
explanation, fine-tuning of the electroweak scale does not really
worsen the problem.  The solution to both hierarchy problems might not
involve natural cancellations, but follow from a completely different
reasoning, such as the idea that galaxy and star formation, chemistry
and biology, are simply impossible without these scales having the
values found in our Universe~\cite{Weinberg:1987dv}.  In the vast
landscape of possible string theory vacua, we may find ourselves in
the observed ground state for exactly these
reasons~\cite{Susskind:2004uv}.\medskip

Supersymmetry has other merits: $R$-parity provides a natural
dark-matter candidate with about the right properties.  Grand
unification is achieved by the quantum corrections due to the gauginos
and Higgsinos.  However, supersymmetry has problems as well: naturalness
is again in conflict with experiment, since the non-observation of
light Higgs bosons and gauginos at LEP requires large, somewhat
fine-tuned, soft-breaking parameters.
Large flavour-changing neutral current~(FCNC) effects due to
sfermion exchange are generically expected but not observed, a problem
that only has a natural solution in gauge-mediated models.
It is also possible that dimension-five
operators at the GUT scale could mediate proton decay with an
unacceptable rate.\medskip

Giving up naturalness of the electroweak scale, the SpS scenario
solves the problems without sacrificing
the merits.  If all sfermions are heavy, it is well known that the pattern of grand
unification is unchanged, since they form complete SU(5)
representations~\cite{Dawson:1979zq}. The Higgs bosons are also expected to be heavy,
but by fine-tuning the $B$ term in the Higgs potential the one Higgs
doublet of the Standard Model can be made light.  This modification of the MSSM
spectrum does not necessarily affect the mass parameters of gauginos
and Higgsinos, which can be protected by the combination of $R$ symmetry and
Peccei--Quinn symmetries.  Hence, a TeV-scale LSP is possible (albeit
not guaranteed).  In the absence of light sfermions, the FCNC and
proton-decay problems completely disappear.

The low-energy effective theory is particularly simple.  In addition
to the Standard Model spectrum including the Higgs boson, the only extra
particles are the four neutralinos, two charginos and a gluino.
Since all squarks are very heavy the gluino is long-lived.
Renormalization group running without sfermions and heavy Higgses
lifts the light Higgs mass considerably above the LEP limit, solving
another problem of the MSSM.  Still, the Higgs boson is expected to be
lighter than about
$200\GeV$~\cite{Arkani-Hamed:2004fb,Giudice:2004tc}.  Apart from
this Higgs mass bound, the only trace of supersymmetry would be the
mutual interactions of Higgses, gauginos and Higgsinos, \ie  
the chargino and neutralino Yukawa couplings.  These
couplings are determined by the gauge couplings at the matching scale
$\sms$, where the scalars are integrated out.  Renormalization group
running yields corrections of the order of 10--20\%
for these couplings~\cite{Giudice:2004tc,Arvanitaki:2004eu}.\smallskip

At the LHC, the experimental challenge would be the observation and
classification of the $R$-hadrons.  In addition, we would like to search for
direct production of the charginos and neutralinos, to identify their gaugino
and Higgsino components, and to measure their Higgs Yukawa couplings.  The
absence of scalar states can be checked by constraining contact
terms. Obviously, these precision
measurements are a perfect task for a high-luminosity linear collider.


\section{Renormalization group evolution}

\begin{figure}[t]
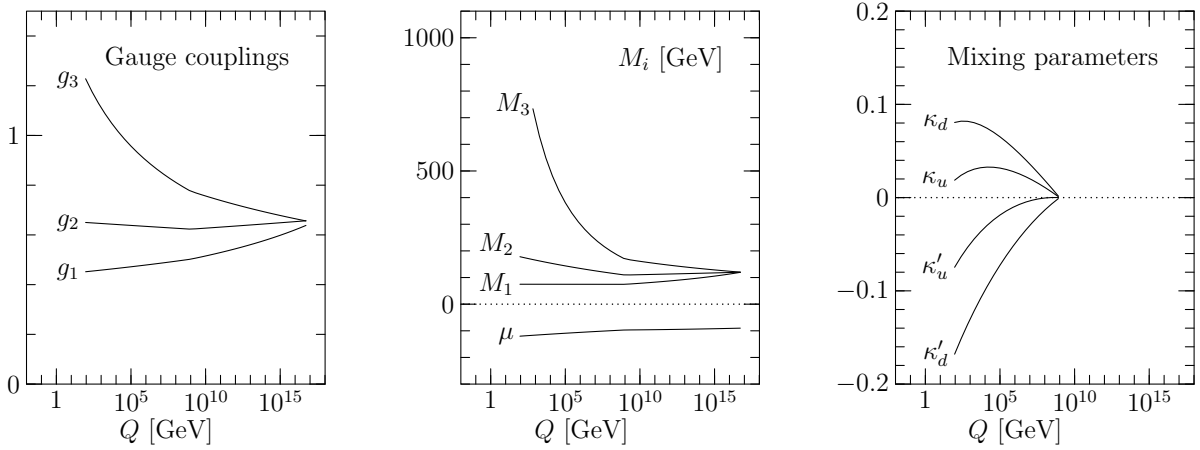

\begin{center}
\includegraphics[width=4.0cm]{rge-plots.1} \hspace{15mm}
\includegraphics[width=4.0cm]{rge-plots.2} \hspace{15mm}
\includegraphics[width=4.0cm]{rge-plots.3}
\end{center}
\vspace{3mm}
\caption{\label{fig:RGE} Renormalization group flow of the gauge
couplings (left), the gaugino--Higgsino mass parameters (centre), and
the anomalous gaugino--Higgsino mixing parameters defined in
eq.(\ref{eq:sps1_kappa}) (right). All curves are based on our reference point 
eq.(\ref{eq:sps1}).}
\end{figure}

In supergravity-inspired SpS models, grand unification relates the
bino, wino and gluino mass parameters $M_{1,2,3}$ at a high scale.
The constraint that the LSP should not overclose the Universe requires
$\mu$ to be not much larger than $M_1$ or
$M_2$~\cite{Giudice:2004tc,Pierce:2004mk}.  At low energies, the SpS
renormalization group enhances the splitting between $M_1$ and $M_2$
on one side and $M_3$ on the other side with respect to the MSSM. The
gluino will be heavy in comparison to the neutralinos and
charginos. All neutralinos and charginos are strongly mixed, because
the $\mu$ parameter should be chosen small at the high scale and then stays
small after renormalization group running.

Assuming gaugino mass unification and a small Higgsino mass parameter,
we start from the following model parameters at the grand unification
scale $\mGUT=6\times 10^{16}\GeV$:
\begin{align}
  M_1(\mGUT) = M_2(\mGUT) &= M_3(\mGUT) = 120\GeV \notag \\
  \mu(\mGUT) &= -90\GeV \notag \\
  \tan\beta  &= 4
\label{eq:init}
\end{align}
For the SUSY-breaking scale we choose $\sms = 10^9\GeV$~\footnote{We stress 
that the phenomenology for this particular choice of $\sms$ is identical 
to the case $\sms=\mGUT$, as will become obvious during the analysis.}. In the
effective theory approach this is the intermediate matching scale
where the scalars are integrated out. Figure~\ref{fig:RGE} displays the
solutions of the renormalization group
equations~\cite{Ferreira:1996ug} following the Appendix of
Ref.~\cite{Giudice:2004tc} with the input parameters set in
eq.(\ref{eq:init}). At the low scale $Q=m_Z$, we extract the mass
parameters:
\begin{align}
  M_1(Q=m_Z) &= 74.8\GeV & \qquad \qquad 
  M_3^{\overline{\text{DR}}}(Q=1 \TeV) &= 690.1\GeV \notag \\
  M_2(Q=m_Z) &= 178.1\GeV  & \qquad \qquad
  \mu(Q=m_Z) &= -120.1\GeV
\label{eq:sps1}
\end{align}
The resulting physical gaugino and Higgsino masses are:
\begin{align}
  m_{\tilde\chi^0_1} &= 71.1\GeV,   & m_{\tilde\chi^+_1} &= 114.7\GeV, \notag \\
  m_{\tilde\chi^0_2} &= 109.9\GeV,  & m_{\tilde\chi^+_2} &= 215.7\GeV, \notag \\
  m_{\tilde\chi^0_3} &= 141.7\GeV, \notag \\
  m_{\tilde\chi^0_4} &= 213.7\GeV,  & m_{\tilde g} &= 807\GeV
\label{eq:sps1_masses}
\end{align}
These mass values satisfy the LEP constraints.  The neutralinos
$\tilde\chi^0_{1,2,3,4}$ are predominantly bino, Higgsino, Higgsino,
and wino, respectively.  The Higgsino content of the lightest
neutralino is $h_f=0.2$, so the dark-matter
condition~\cite{Pierce:2004mk} is satisfied.  To our given order the
Higgs mass is $m_H=150\GeV$, but as usually it will receive sizeable
radiative corrections~\cite{Heinemeyer:2004ms}.

Because we integrate out the heavy scalars, the neutralino and
chargino Yukawa couplings deviate from their usual MSSM prediction,
parametrized by four anomalous Yukawa couplings
$\kappa$~\cite{Arvanitaki:2004eu}. We can extract their weak-scale values
from Fig.~\ref{fig:RGE}:
\begin{align}
  \frac{\tilde g_u}{g\sin\beta}   &\equiv 1 + \kappa_u
                                   = 1 + 0.018
 &\frac{\tilde g_d}{g\cos\beta}   &\equiv 1 + \kappa_d
                                   = 1 + 0.081 \notag \\ 
  \frac{\tilde g'_u}{g'\sin\beta} &\equiv 1 + \kappa'_u
                                   = 1 - 0.075
 &\frac{\tilde g'_d}{g'\cos\beta} &\equiv 1 + \kappa'_d
                                   = 1 - 0.17
\label{eq:sps1_kappa}
\end{align}
Note that these are the leading logarithmic renormalization group
effects, which should be supplemented by the complete one-loop
corrections to the neutralino and chargino mixing
matrices~\cite{Fritzsche:2002bi}.

\section{Signals at the LHC}

\begin{table}[b]
\begin{center} \begin{tabular}{|c|r||c|r||c|r||c|r|}
\hline
   & $\sigma[\fb]$ &    & $\sigma[\fb]$ &    & $\sigma[\fb]$ & & $\sigma[\fb]$ \\
\hline
$\tilde g \tilde g$ & 1710   &&&&&&                                                   \\ \hline 
$\cme\cpe$ & 2910  & $\cme\cpz$ & 73.7  & $\cpe\cmz$ &  73.7 & $\cpz\cmz$ &  604  \\ \hline
$\nne\nne$ & 49.4  & $\nne\nnz$ & 49.7  & $\nne\nnd$ &  409  & $\nne\nnv$ & 0.06  \\
           &       & $\nnz\nnz$ &  5.0  & $\nnz\nnd$ &  876  & $\nnz\nnv$ &  3.7  \\
           &       &            &       & $\nnd\nnd$ &  1.4  & $\nnd\nnv$ & 69.6  \\ 
           &       &            &       &            &       & $\nnv\nnv$ &  1.0  \\ \hline
$\cme\nne$ &  584  & $\cme\nnz$ & 1780  & $\cme\nnd$ &  789  & $\cme\nnv$ & 78.8  \\
$\cpe\nne$ &  914  & $\cpe\nnz$ & 2870  & $\cpe\nnd$ & 1310  & $\cpe\nnv$ &  138  \\
$\cmz\nne$ &  2.7  & $\cmz\nnz$ & 55.9  & $\cmz\nnd$ & 66.6  & $\cmz\nnv$ &  430  \\
$\cpz\nne$ &  4.5  & $\cpz\nnz$ & 97.7  & $\cpz\nnd$ &  119  & $\cpz\nnv$ &  798  \\ \hline
\end{tabular} \end{center}
\caption{\label{tab:lhc} NLO production cross sections at the
LHC~\cite{Beenakker:1996ch}. The masses and mixing matrices are
fixed by the reference point in eq.(\ref{eq:sps1}).}
\end{table}

The production cross section of gluinos and of charginos and
neutralinos at the LHC are known to NLO~\cite{Beenakker:1996ch}. In
SpS models these cross sections depend only on the gluino mass and the
chargino and neutralino masses and mixings, respectively, the latter
being determined by the gaugino mass parameters $M_1$ and $M_2$ and
the Higgsino mass~$\mu$. In Table~\ref{tab:lhc} we list the LHC
production cross sections for our example parameter point in
eq.(\ref{eq:sps1_masses}). For neutralino pairs we can understand the
simple pattern, since in the heavy squark limit, neutralino production
only proceeds through a Drell--Yan $s$-channel $Z$ boson. The
neutralinos with a large Higgsino fraction are $\tilde{\chi}^0_{2,3}$,
which makes the $\nnz\nnd$ production dominant. The pair production of
either of these two states is suppressed because the couplings of the
$Z$ boson to the two Higgsino states cancel each other in the
superposition. Diagonal chargino pairs are produced at a comparably
large rate because of the $s$-channel photon exchange. 

The charginos are strongly mixed, but the ligher $\tilde{\chi}^\pm_1$
has a larger Higgsino and the heavier $\tilde{\chi}^\pm_2$ has a
larger wino fraction.  In mixed chargino and neutralino production,
the $s$-channel $W$ boson couples to either a
$\tilde{H}^0\tilde{H}^\pm$ or to a $\tilde{W}^0\tilde{W}^\pm$
combination. Because of the composition of the neutralinos, the
production cross sections for $\tilde{\chi}^\pm_1 \nnz$,
$\tilde{\chi}^\pm_1 \nnd$, and $\tilde{\chi}^\pm_2 \nnv$ are dominant
(The combination of the lightest chargino and neutralino benefits from
their small masses and the sizeable mixing). The final states with a
positive charge have a typically twice as large cross section as the
final states with negative charge, due to the valence quark
decomposition of the initial-state proton.

Strategies to discover MSSM particles and measure their masses at the
LHC usually rely on the production of squarks and gluinos and
subsequent cascade decays to the weakly interacting superpartners. In
SpS scenarios this is not possible. Instead, we have to look for 
direct Drell--Yan-like production channels, which are plagued by overwhelming
$W$ and $Z$ production backgrounds. In particular the trilepton
signature $pp\to\nnz\cpe$  becomes considerably harder to observe if the decay $\nnz
\to \ell^+ \ell^- \nne$ does not involve an intermediate slepton. Our
SpS parameter point with the masses given in eq.(\ref{eq:sps1_masses})
does not allow for the decay $\nnz \to H \nne$, but for heavier
particles this decay might be promising to look for gauginos and
Higgsinos at the LHC.  Note that the associated production of
charginos and neutralinos with a gluino is mediated by a $t$-channel
squark exchange, and therefore is suppressed.

\subsection{Gluino decays}

\begin{figure}[b]
\begin{center}
\includegraphics{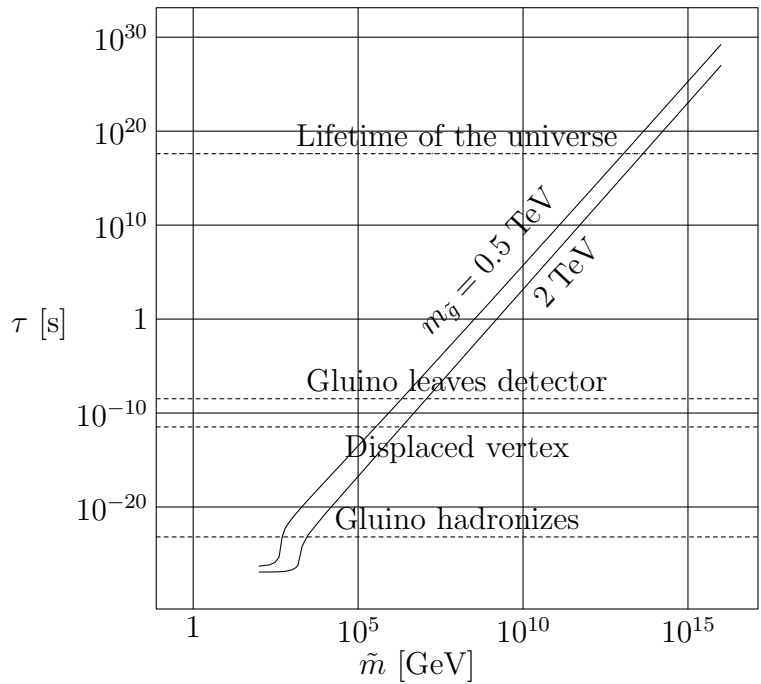}
\end{center}
\vspace{3mm}
\caption{\label{fig:gluino} Gluino lifetime~\cite{Muhlleitner:2003vg}
  as a function of the common scalar mass $\sms$.}
\end{figure}

Unless we have a priori knowledge about the sfermion scale $\sms$,
the gluino lifetime is undetermined. Figure~\ref{fig:gluino} compares
this scale with other relevant scales of particle physics.  Once
$\sms\gtrsim 10^{3}\GeV$, the gluino hadronizes before decaying.
For $\sms>10^{6}\GeV$, weak decays of heavy-flavoured $R$-hadrons
start to play a role, and the gluino travels a macroscopic distance.
If $\sms>10^{7}\GeV$, strange $R$-hadrons can also decay weakly, and
gluinos typically leave the detector undecayed or are stopped in the
material.  For even higher scales, $\sms>10^{9}\GeV$, $R$-hadrons
could become cosmologically relevant, since they affect
nucleosynthesis if their abundance in the early Universe is
sufficiently high~\cite{Arkani-Hamed:2004fb,Giudice:2004tc}.  Finally,
$\sms>10^{13}\GeV$ is equivalent to a stable gluino since its
lifetime is longer than the age of the Universe.
With the given value of the intermediate scale $\sms=10^9\GeV$ and the
weak-scale parameters of eq.(\ref{eq:sps1}), the gluino width is of the order of
\begin{equation}
  \Gamma_{\tilde g} \approx 1.0\times 10^{-25}\GeV,
  \qquad \ie \qquad
  \tau_{\tilde g} \approx 6.5\;\s.
\end{equation}
The precise value can be computed only if the detailed squark
spectrum is known; the above numbers correspond to universal scalar
masses and no mixing.

If gluino decays can be observed, their analysis yields information
about physics at the scale $\sms$ and thus allows us to draw
conclusions about the mechanism of supersymmetry breaking.  In a
standard MSSM scenario with heavy scalars, the gluino will experience
a three-body decay $\tilde g \to q\bar q\tilde\chi^0$ or 
$\tilde g \to q\bar q'\tilde\chi^\pm$.
The $\tilde\chi$'s are predominantly gaugino for light quarks in the
final state.  In the charged decay, the flavour mixing is governed by
the standard CKM matrix.  A loop-induced decay
$\tilde g \to g\tilde\chi^0$
 is also possible and has a rate comparable to the tree-level three-body decays.  This
decay mainly proceeds via a top/stop loop, and the neutralino is
predominantly Higgsino.\smallskip

In the usual MSSM scenarios squarks are degenerate, so these
decays are flavour-diagonal modulo CKM effects and left--right
squark mixing in the third generation. The situation is different in
SpS:  due to the absence of FCNC constraints, arbitrary
sfermion mass patterns are allowed once the scalar mass $\sms$
exceeds a value of order $10^5\GeV$.  On the other hand, the
left--right sfermion mixing angles vanish in SpS since
the off-diagonal elements of the mixing matrices are suppressed by
$v/\sms$.
Therefore, the flavour decomposition of gluino decays mirrors the
sfermion mass hierarchy at the matching scale $\sms$.  The ratio of
branching ratios $\tilde g\to q\bar q\tilde\chi^0$ and $\tilde g\to
Q\bar Q\tilde\chi^0$ is given by $(m_{\tilde Q}/m_{\tilde q})^4$, so
even a weak hierarchy will be greatly enhanced in the branching
ratios.  
If the decays of long-lived gluinos can be observed, it
is important to identify flavour, even though the conditions of
flavour-tagging are non-standard if the decay does not occur near the
interaction point.

\subsection{\boldmath$R$-hadrons}

$R$-hadrons have been discussed early in supersymmetry
phenomenology~\cite{Farrar:1978xj}.  The spectrum of light-flavoured
$R$-hadrons can be computed using a bag model~\cite{Chanowitz:1983ci}
or lattice calculations~\cite{Foster:1998wu}.
The gluino is a colour octet,
therefore a colour-singlet hadron can be made by adding a
quark--antiquark pair coupled as an octet (in SU(3),
$3 \otimes \bar 3 = 8 \oplus 1$), or
by three quarks coupled as an octet, which is possible in two ways
($3 \otimes 3 \otimes 3 = 10 \oplus 8 \oplus 8' \oplus 1$).
Furthermore, the gluino colour can be neutralized by adding a single
constituent gluon (or another gluino, for that matter).  These neutral
states are collectively denoted by $R_g$.  In any case, for a heavy
gluino the mass differences of the various $R$ states are small
with respect to the overall mass.  This situation is described by
heavy-quark effective theory, where the gluino acts as a static colour
source, unaffected by the dynamics of the quark--gluon cloud around it~\cite{Neubert:1993mb}.\medskip

The $\tilde g\bar qq$ hadrons are similar to ordinary mesons, and they
may be labelled in an analogous way: $R_\pi,R_\rho,R_K,\ldots$
The total spin is fermionic ($1/2$ or $3/2$), but this does not
affect the dynamics because the gluino spin decouples from the
surrounding cloud and a meson description is therefore appropriate.  The higher
excitations rapidly decay into the lowest excitations. Considering the
ground states, we note that the $R_\pi$ hadrons are not Goldstone
bosons, so they are not particularly light.  The numerical estimates
in~\cite{Foster:1998wu} indicate that the $R_\rho$ states are slightly
lighter, and the lowest $R_g$ is close to it.  However, all mass differences
are expected to be less than $100\;\MeV$, so that all these ground states
are stable with respect to the strong interaction. As long
as the gluino decays at all, there is no reason for the
lowest state to be neutral, so after weak decays the final state
of the $R$-hadron decay chain could be either a neutral $R$ or, say,
the $R_\rho^\pm$. In analogy to the mixing of the $\rho^0$ and the
photon, we expect mixing of $R_\rho^0$ and $R_g$, so there may be
significant isospin-breaking effects.

The $R$-baryon spectrum differs considerably from ordinary baryons by
the different colour- and flavour-coupling schemes allowed.
Their spectrum can be estimated using bag models~\cite{Buccella:1985cs}.
However, in the process of fragmentation,
baryon formation is less likely than meson formation, a feature that
should persist in the present situation.\medskip

If the gluino production rate is sufficiently large,
heavy-flavoured $R$-hadrons can be produced.  These are
interesting objects, because their weak decays may provide distinctive
signatures of SpS.  Let us consider the $R_B^-=\tilde gb\bar u$.  In
the field of the static source $\tilde g$, the $b$ quark will tightly
bind to it, since $m_b \gg \Lambda_{\rm QCD}$.  This system is
approximately described by the same perturbative potential as
describes the lowest-lying $\Upsilon$ states.  The difference is that
the physical $b$ mass should be used instead of the reduced mass
$m_b/2$, and therefore $\alpha_s$ should be evaluated at a slightly
higher scale. Moreover, the prefactor $4/3$, which is appropriate for
$3 \otimes \bar 3 \to 1$ coupling we must 
replace by $3/2$, which corresponds to
$8 \otimes 3 \to 3$. (Note that the triplet
channel is the most attractive one in the coupling of an octet and a
singlet.)  Assuming that the systems are Coulombic, we can
estimate the $R_b$ binding energy:
\begin{equation}
  E(R_b) \approx \frac{9}{4}\,\frac{\alpha(m_b)}{\alpha(m_b/2)}\,E(\Upsilon)
\end{equation}
If we take $m_{\Upsilon(1S)}-2m_B=-1\GeV$ as the $\Upsilon$ binding
energy, we obtain $E(R_b)\approx -2\GeV$.  Clearly, this estimate can
be refined by looking at the potential in more detail.

The gluino--$b$ system forms a colour-triplet nucleus, which is
surrounded by the light-quark cloud.  Heavy-quark symmetry tells us
that the dynamics of this cloud is similar to the dynamics of an
ordinary $B$-meson.  More precisely, the orbital part of the
Hamiltonian is the HQET Hamiltonian in the
extreme heavy-mass limit.  The spin part is identical to the $B$-meson
Hamiltonian, since only the $b$-quark spin couples to the light cloud,
suppressed by $\Lambda_{\rm QCD}/m_b$, while the gluino spin is
irrelevant.  Thus, $B$-meson data can be exploited to determine many
of the properties of these states.

\subsection{Long-lived gluinos at the LHC}

  The phenomenology of SpS models at the LHC is very dependent on the 
  lifetime of the gluino. If this is smaller than the hadronization 
  time scale, the signals will be the usual signals for supersymmetry~\cite{atlastdrII}.
  If the gluino hadronizes, but the lifetime of the $R$-hadrons
  produced is short enough that they decay inside the detector, 
  we might see additional vertices from the $R$-hadron decay.
  Because the phenomenology of 
  these scenarios has been extensively considered in the
  literature, we will only consider the long-lived gluino here.
  For a stable $R$-hadron we investigate two types of signals: 
\begin{enumerate}
\item The production of a stable, charged, $R$-hadron will give a signal 
      much like the
      production of a stable charged weakly-interacting particle~\cite{Allanach:2001sd,Hinchliffe:1998ys}.
      This signal consists of an object that looks like a muon but arrives
      at the muon chambers significantly later than a muon owing to its large mass.
      However, the situation will be more complicated than those considered in
      Refs.~\cite{Allanach:2001sd,Hinchliffe:1998ys},
         as the $R$-hadron will interact more in the
      detector, losing more energy.
\item While for stable neutral $R$-hadrons there will be some energy loss 
      in the detector, 
      there will be a missing transverse energy signal due to the escape of
      the $R$-hadrons. As leptons are unlikely to be produced in this process
      the signal will  be the classic SUSY jets with missing transverse energy
      signature.
\end{enumerate} 
  There is also the possibility of signals involving the production and semileptonic 
  decay of  $R$-hadrons containing a heavy, \ie bottom or charm, quark.
  For a gluino, which is stable on collider time scales,
  the phenomenology of the model depends on 
  the cross section, which is controlled by the gluino 
  mass, the ratio of stable charged to neutral $R$-hadrons produced and, for the 
  decay of $R$-hadrons containing a heavy quark, the number of these hadrons produced.
  \medskip

  To study the production of $R$-hadrons at the LHC we have to
  model the hadronization of the gluino. Our simulations use 
  the HERWIG Monte 
  Carlo event generator~\cite{Corcella:2000bw}, which in turn uses the
  cluster hadronization
  model~\cite{Webber:1983if}. In HERWIG,
  the gluons left at the end of the perturbative evolution in
  the QCD parton shower are non-perturbatively split into quark--antiquark pairs.
  In the large-$N_C$ limit, the quarks and antiquarks can be
  uniquely formed into colour-singlet clusters that carry mesonic quantum numbers.
  Preconfinement ensures that these clusters have a mass spectrum that peaks at low
  values and falls off rapidly at higher masses.\medskip

  These clusters are assumed to be a superposition of the known hadron resonances
  and decay into two hadrons.
  To illustrate the decay we consider a cluster containing
  a quark $q_i$ and an antiquark $\bar{q}_j$ ($i,j$ are flavour indices). First, a 
  quark--antiquark pair of flavour $k$ is produced from the vacuum with probability
  $P_k$.\footnote{The probabilities $P_k$ are parameters of the model. They are normally
                 set so that the probabilities are equal for the
                 light (up, down and strange) quarks and equal to zero for the 
                 heavy (charm and bottom) quarks.}
  This specifies the flavours of the two produced mesons, $(q_i\bar{q}_k)$ and $(q_k\bar{q}_j)$.
  The type of meson is randomly chosen from the available mesons with the correct
  flavours. A weight
\begin{equation}
    W = (2S_{q_i\bar{q}_k}+1) \; (2S_{q_k\bar{q}_j}+1) \;
        \Phi\left[C\to(q_i\bar{q}_k),(q_k\bar{q}_j)\right],
\end{equation}
  where $S_{q_i\bar{q}_k}$ and $S_{q_k\bar{q}_j}$ are the spins of the mesons selected
  and $\Phi\left[C\to(q_i\bar{q}_k),(q_k\bar{q}_j)\right]$ is the two-body phase-space
  weight for the
  decay of the cluster, is then calculated.
  A decay is accepted if $W\geq\mathcal{R}W_{\rm max}$
  where $\mathcal{R}$ is a random number between 0~and~1 and $W_{\rm max}$ is the 
  maximum possible weight. If the decay is rejected the procedure is repeated and a 
  new quark--antiquark pair and types of mesons are selected.

  If we include a gluino, which is stable on the hadronization time scale,
  we will have a cluster containing a gluino in addition to the quark $q_i$ 
  and the antiquark $\bar{q}_j$.
  The simplest approach would be to select a quark--antiquark pair $(q_k\bar{q_k})$
  as before, and randomly select either $(q_i\tilde{g}\bar{q}_k)$ and $(q_k\bar{q}_j)$
  or $(q_i\bar{q}_k)$ and $(q_k\tilde{g}\bar{q}_j)$ as the flavours for the mesons.
  However this would forbid direct production of the $R_g$ hadrons and in 
  particular the lightest $R$-hadron state $R_g^0$. There is no obvious mechanism for
  the production of the $R_g$ states in the cluster model; we therefore chose to
  model $R_g$ production by including the decay of a cluster containing a gluino
  to $R_g$ and the lightest meson, with quark flavours $(q_i\bar{q_j})$, 
  in addition to the normal cluster decays. This decay 
  occurs with a probability $P_{R_g}$. The parameter $P_{R_g}$ will generally act
  as a parametrization of how many of the $R$ mesons in the detector are
  neutral and how many are charged. This fraction determines the relative success of
  the two search strategies listed above.\medskip

  In order to simplify the simulation we only include the lightest
  $R$-hadron with a given quark composition and do not include the $R$-baryons. 
  The lightest $R$-hadron is taken to be the lightest $R_g$ state ($R^0_g$)
  with mass $M_{\tilde{g}}+1.43\GeV$~\cite{Karl:1999wq}.
  The lowest-lying $R$-meson is the $R_\rho$ with a mass $M_{R^0_g}+47\;\MeV$~\cite{Foster:1998wu}.
  The masses of the remaining $R$-hadrons are then given by
\begin{equation}
   M_{R_{q_i\bar{q}_j}} = M_{R_\rho} + m_{q_i}+m_{q_j}-2m_{u,d},
\end{equation}
  where $m_{q_i}$ is the constituent mass for the quark of flavour~$i$ and 
  $m_{u,d}$ is the common constituent mass for the up- and down-type quarks.

\begin{table}[t]
\begin{small}
\begin{center}
\begin{tabular}{|c||c|c||c|c|}
\hline
           & \multicolumn{2}{|c||}{$M_{\tilde{g}}=50\GeV$}
           & \multicolumn{2}{|c|}{$M_{\tilde{g}}=2000\GeV$} \\ \cline{2-5}
$R$-hadron & Number  & Percentage of & Number & Percentage of\\
& per $\rm{fb}^{-1}$& $R$-hadrons &  per $\rm{fb}^{-1}$ & $R$-hadrons\\
\hline
 $R_{\rho^0}$           & $(4.152\pm0.006)\times10^8$   &$28.10\pm0.04$                 
                        &$0.5576\pm0.0007$              &$28.22\pm0.04$                 \\
 $R_{\rho^-}$           & $(2.067\pm0.004)\times10^8$   &$14.00\pm0.03$                 
                        &$0.2788\pm0.0005$              &$14.11\pm0.07$                 \\
 $R_{\rho^+}$           & $(2.076\pm0.004)\times10^8$   &$14.05\pm0.03$                 
                        &$0.2788\pm0.0005$              &$14.11\pm0.07$                 \\
 $R_{K^0}$              & $(1.302\pm0.003)\times10^8$   &$8.81\pm0.02$                  
                        &$0.1730\pm0.0004$              &$8.76\pm  0.02$                \\
 $R_{\bar{K}^0}$        & $(1.291\pm0.003)\times10^8$   &$8.74\pm0.02$                  
                        &$0.1730\pm0.0004$              &$8.76\pm  0.02$                \\
 $R_{K^+}$              & $(1.300\pm0.003)\times10^8$   &$8.80\pm0.02$                  
                        &$0.1728\pm0.0004$              &$8.75\pm  0.02$                \\
 $R_{K^-}$              & $(1.299\pm0.003)\times10^8$   &$8.79\pm0.02$                  
                        &$0.1725\pm0.0004$              &$8.73\pm  0.02$                \\
 $R_{\eta }$            & $(1.286\pm0.003)\times10^8$   &$8.71\pm0.02$                  
                        &$0.1687\pm0.0004$              &$8.54\pm  0.02$                \\
 $R_{D}$                & $(2.1\pm0.7)\times10^4$       &$(14.5\pm2.6)\times10^{-4}$    
                        &$(6.5\pm0.8)\times10^{-5}$     &$(3.2\pm0.4)\times10^{-3}$     \\
 $R_{B}$                & $(7\pm7)\times10^3$           &$(0.5\pm0.5)\times10^{-4}$     
                        &$8.0\pm2.8\times10^{-6}$       &$(0.4\pm  0.2)\times10^{-3}$   \\
 $R_{D_s}$              & $(20\pm4)\times10^4$          &$(14.0\pm2.6)\times10^{-4}$    
                        &$4.7\pm0.7\times10^{-5}$       &$(2.4\pm  0.4)\times10^{-3}$   \\
\hline                  
\end{tabular}   
\end{center}                    
\end{small}
\caption{\label{tab:Rproduction} Production rates of the
         $R$-hadrons. The probability of producing an $R_g^0$ is set
         to zero when producing these numbers. The rates of the
         $R_{q\bar{q}}$ hadrons linearly decrease, and $R_g^0$
         increase, as this probability is increased.}
\end{table}             
                        
  The lightest mesonic $R_\rho$ states are taken to be stable, as is the lightest
  gluonic state $R^0_g$. The heavier mesonic $R$-hadrons decay weakly to the
  appropriate lighter $R$-hadron and an off-shell $W$-boson, which decays
  either leptonically or hadronically. The $R_\phi$ state is too light to 
  have a strong decay and it therefore decays to a pion and the $R_\rho$, in
  analogy with the Standard Model decay $\phi\to\rho\pi$. The decay of mesons
  containing a pair of heavy quarks does not need to be modelled, because these
  are not produced in our approximation.

  Assuming both, the $R_\rho$ and the $R^0_g$ to be stable is based on the 
  observation that their mass difference is smaller than the pion mass. The
  actual ordering of their masses does not play any role in our analysis.
  However, the mass difference could of course be much larger. In that 
  case all $R$-hadrons would decay to one distinctly lightest state. The 
  charge of this final state would in turn decide which of our two search 
  strategies for long-lived gluinos will be successful at the LHC. 
  \medskip
 
\begin{table}[t]
\begin{small}
\begin{center}
\begin{tabular}{|c|c|c|c|}
\hline
Detector system & Radius~[m] & Length~[m] & Number of interaction lengths \\
\hline
Inner detector              & 1.15 & 3.5  & 0.0   \\
Electromagnetic calorimeter & 2.25 & 6.65 & 1.2  \\
Hadronic calorimeter        & 4.25 & 6.65  & 9.5  \\
Support structure           & 10   & 20   & 1.5 \\
\hline                
\end{tabular}
\end{center}
\end{small}
\caption{\label{tab:detector} Size of the detector systems. The size of
         the components and their depth in terms of interaction
         lengths are based on the parameters of the ATLAS
         detector~\cite{atlastdrI}.}
\end{table}

\begin{figure}[t]
\includegraphics[angle=90,width=0.45\textwidth]{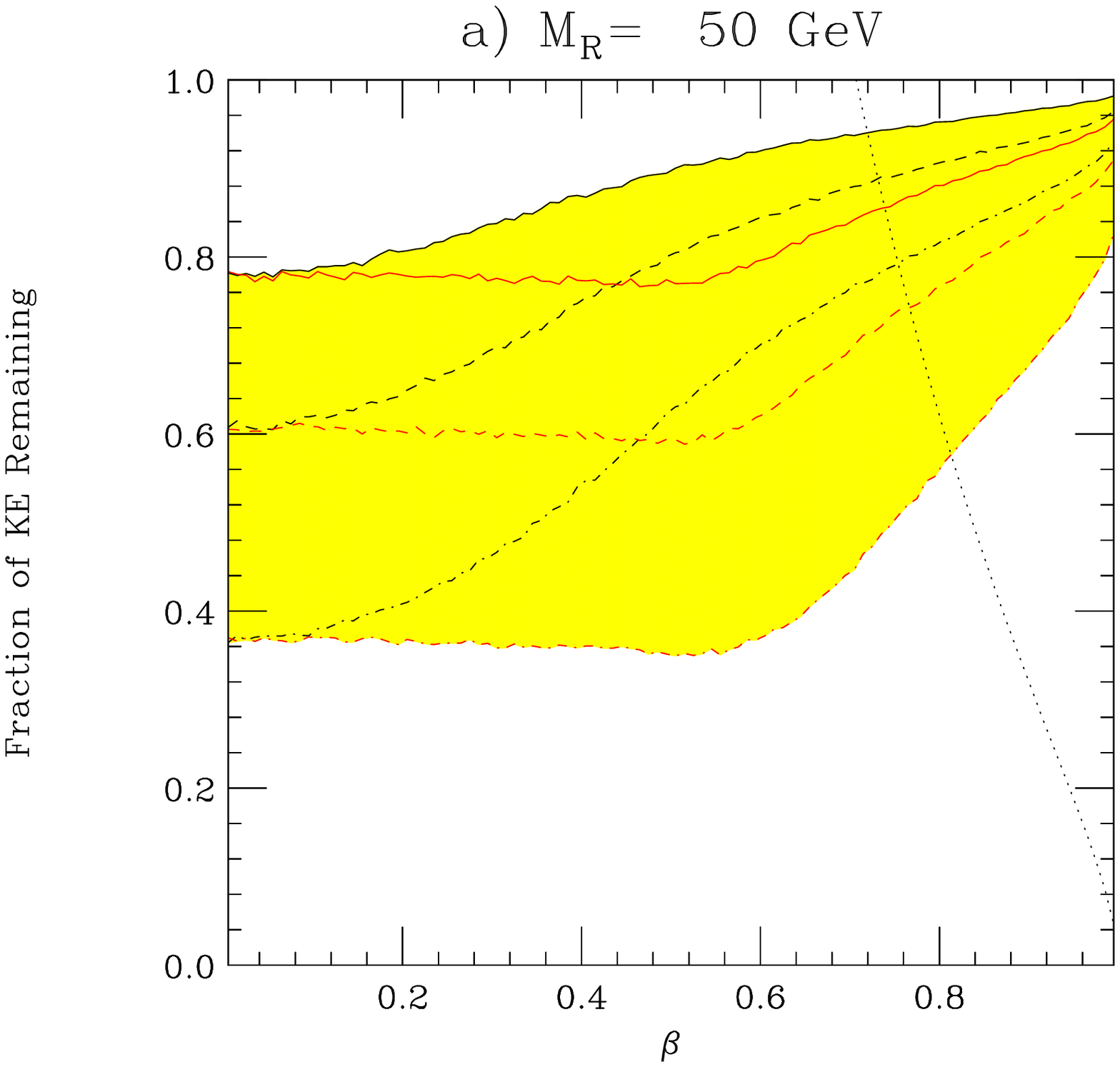}\hfill
\includegraphics[angle=90,width=0.45\textwidth]{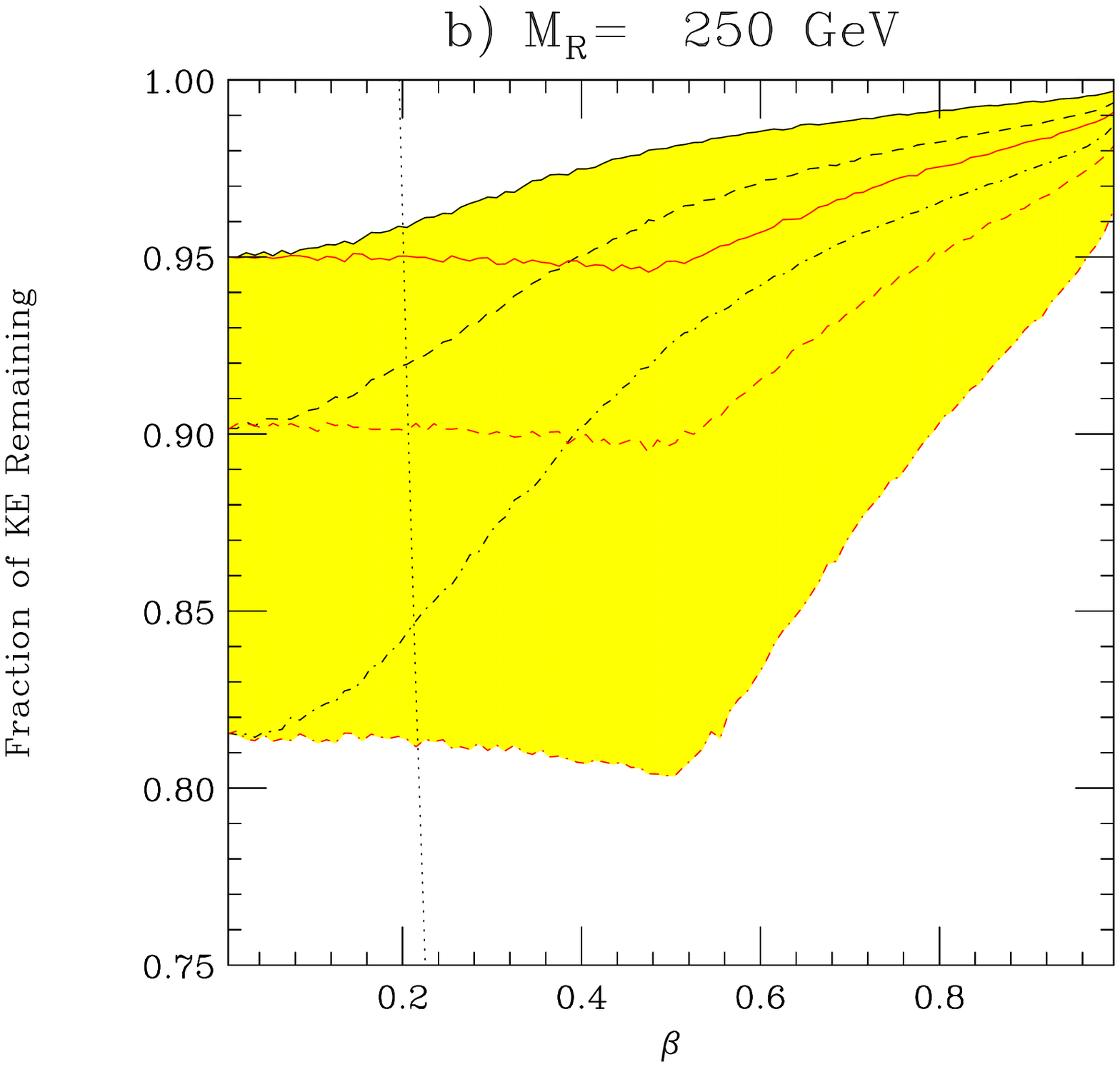}\\
\includegraphics[angle=90,width=0.45\textwidth]{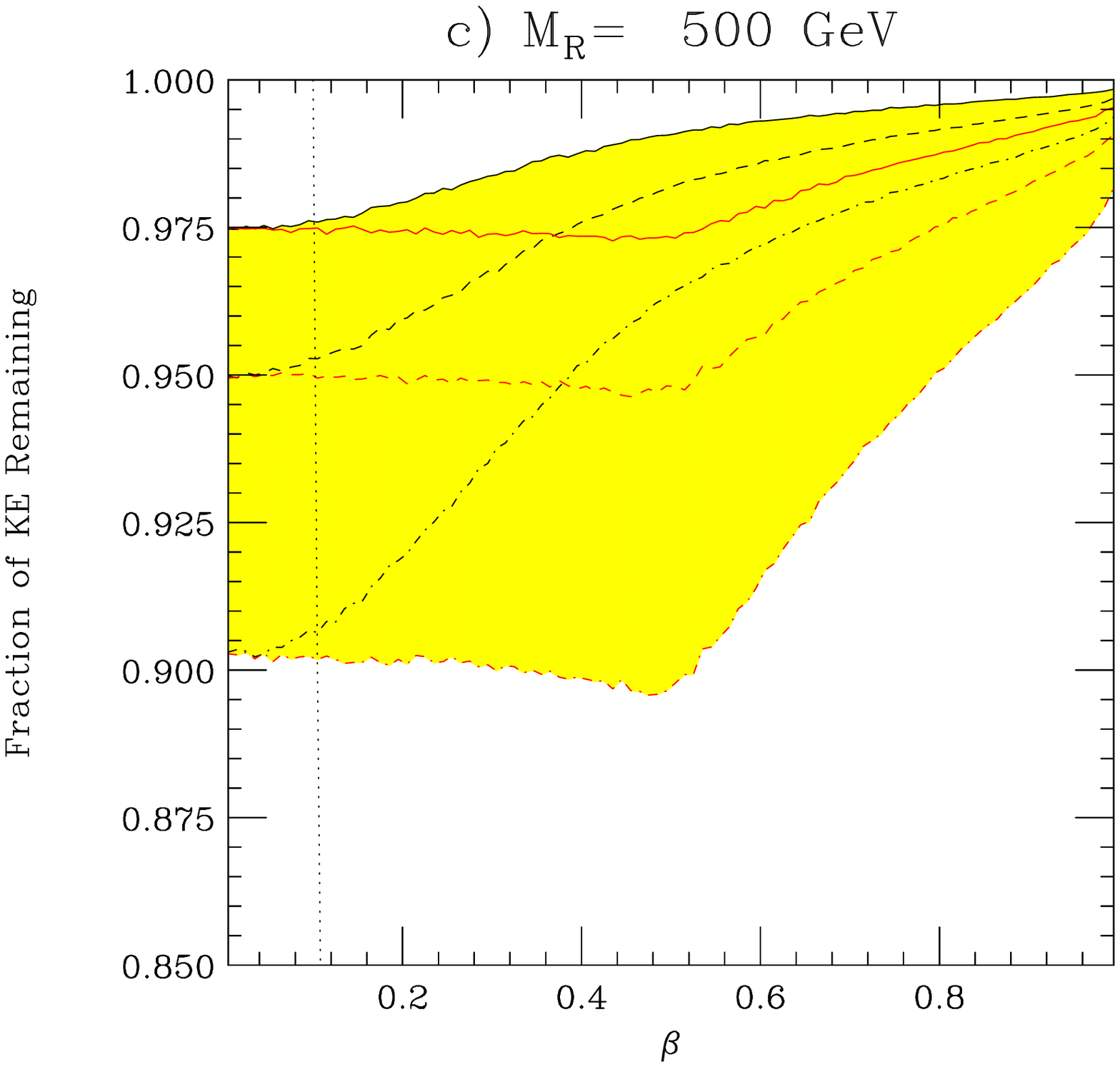}\hfill
\includegraphics[angle=90,width=0.45\textwidth]{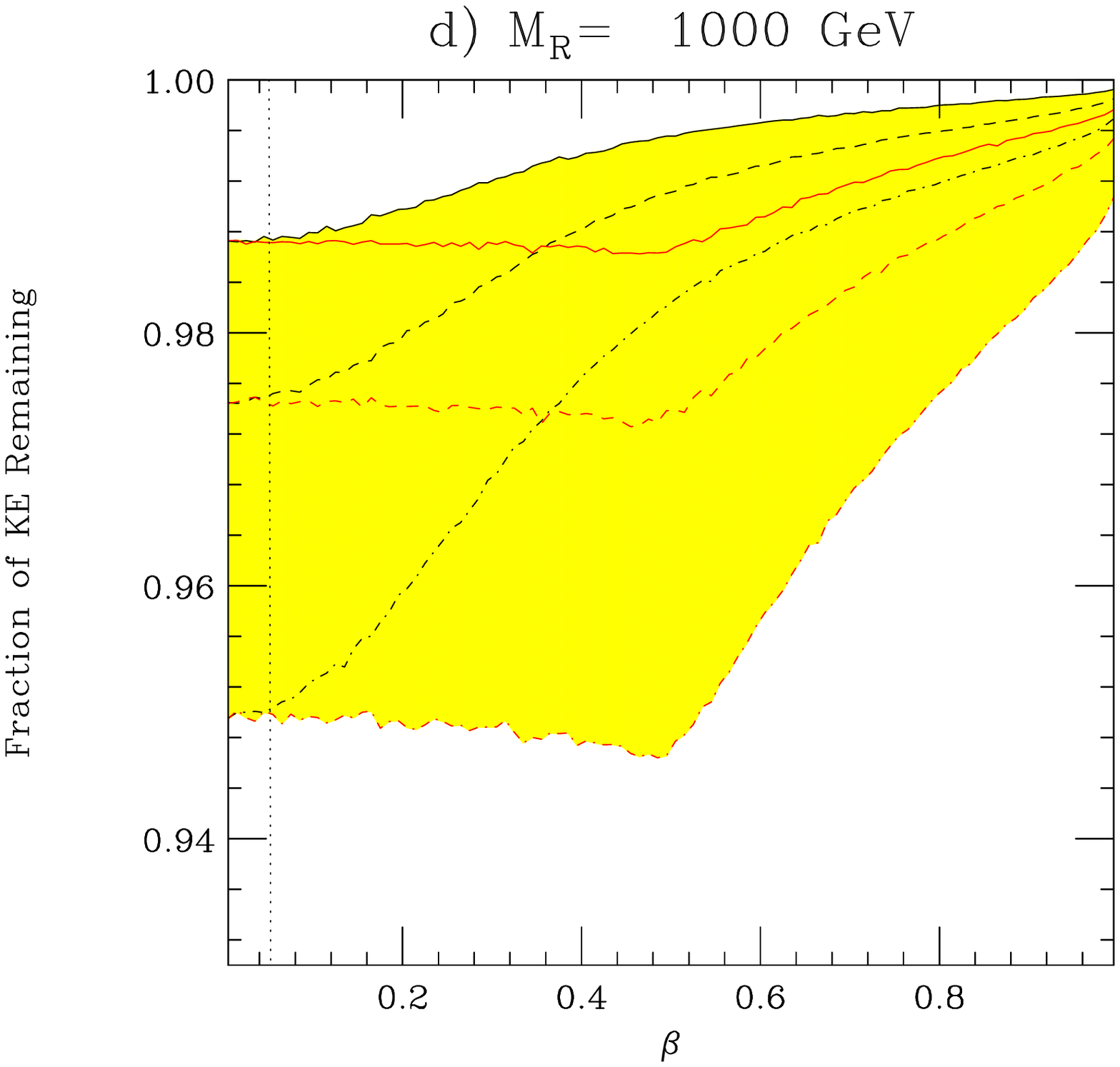}\\
\vspace*{-5mm}
\caption{\label{fig:energyloss}
        Fraction of the kinetic energy remaining when the $R$-hadron enters the muon
        detector. The black lines are for the triple-pomeron form of the $R$-hadron
        nucleon cross section, the red lines are for
        the simple cut-off form~\cite{Baer:1998pg}.
        The interaction length $\lambda_{R}$ is varied between twice the central value (solid),
        the central value (dashed) and half this value (dot-dashed).
        The dotted line shows where the $R$-hadron no longer passes the $p_T$
        cut for the charged $R$-hadron analysis.}
\end{figure}

\begin{figure}[t]
\includegraphics[angle=90,width=0.45\textwidth]{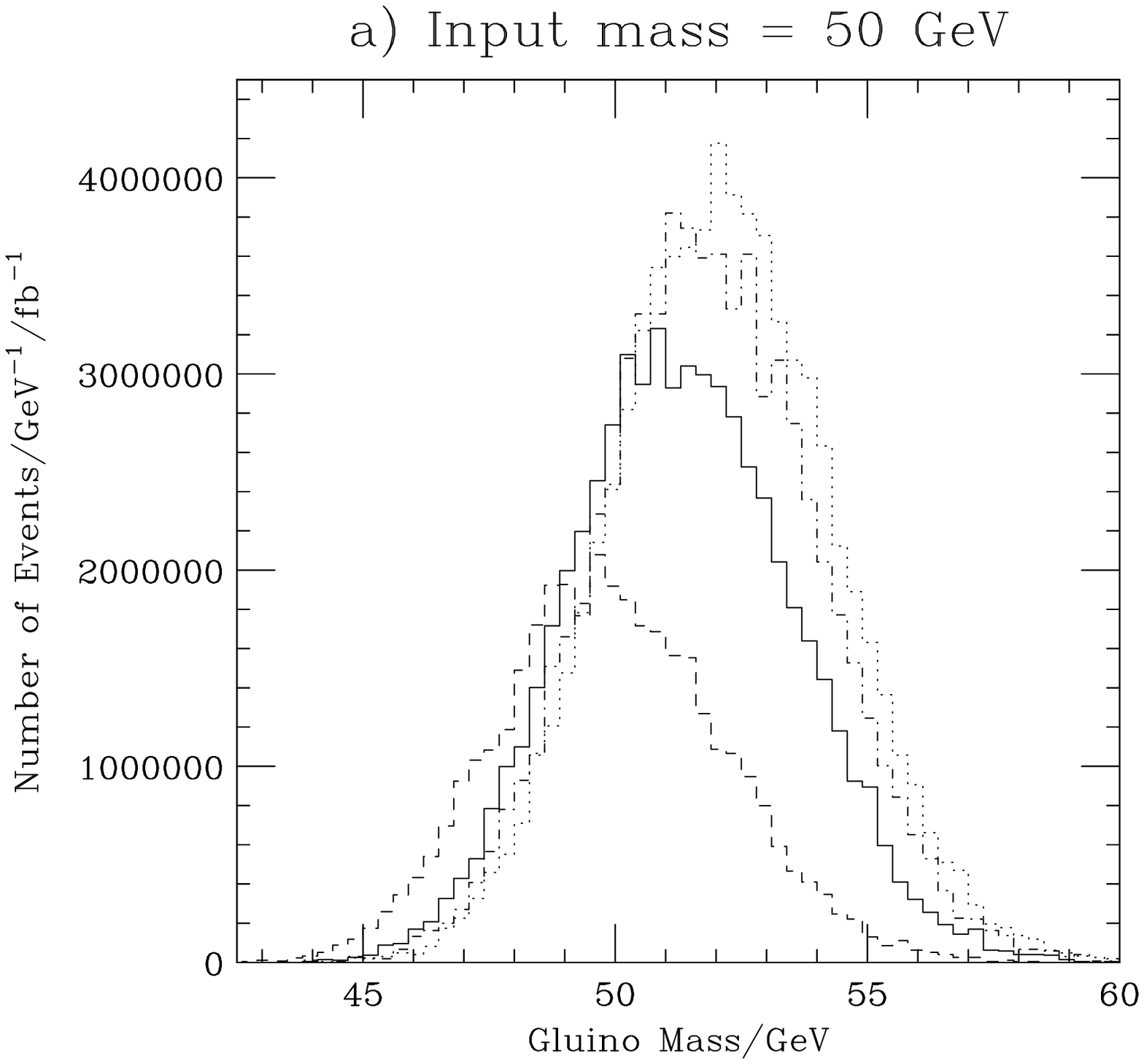}\hfill
\includegraphics[angle=90,width=0.45\textwidth]{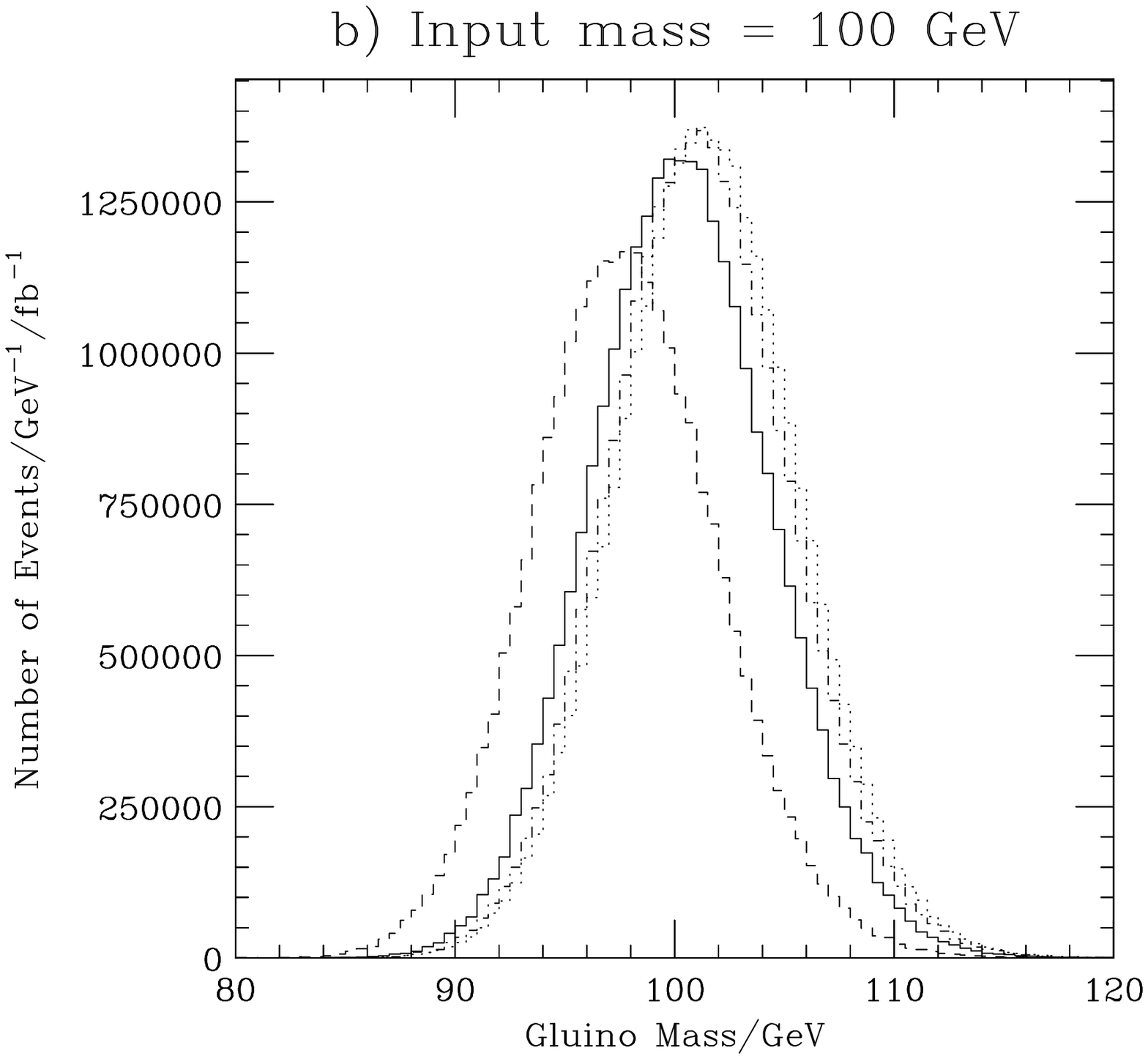}\\
\includegraphics[angle=90,width=0.45\textwidth]{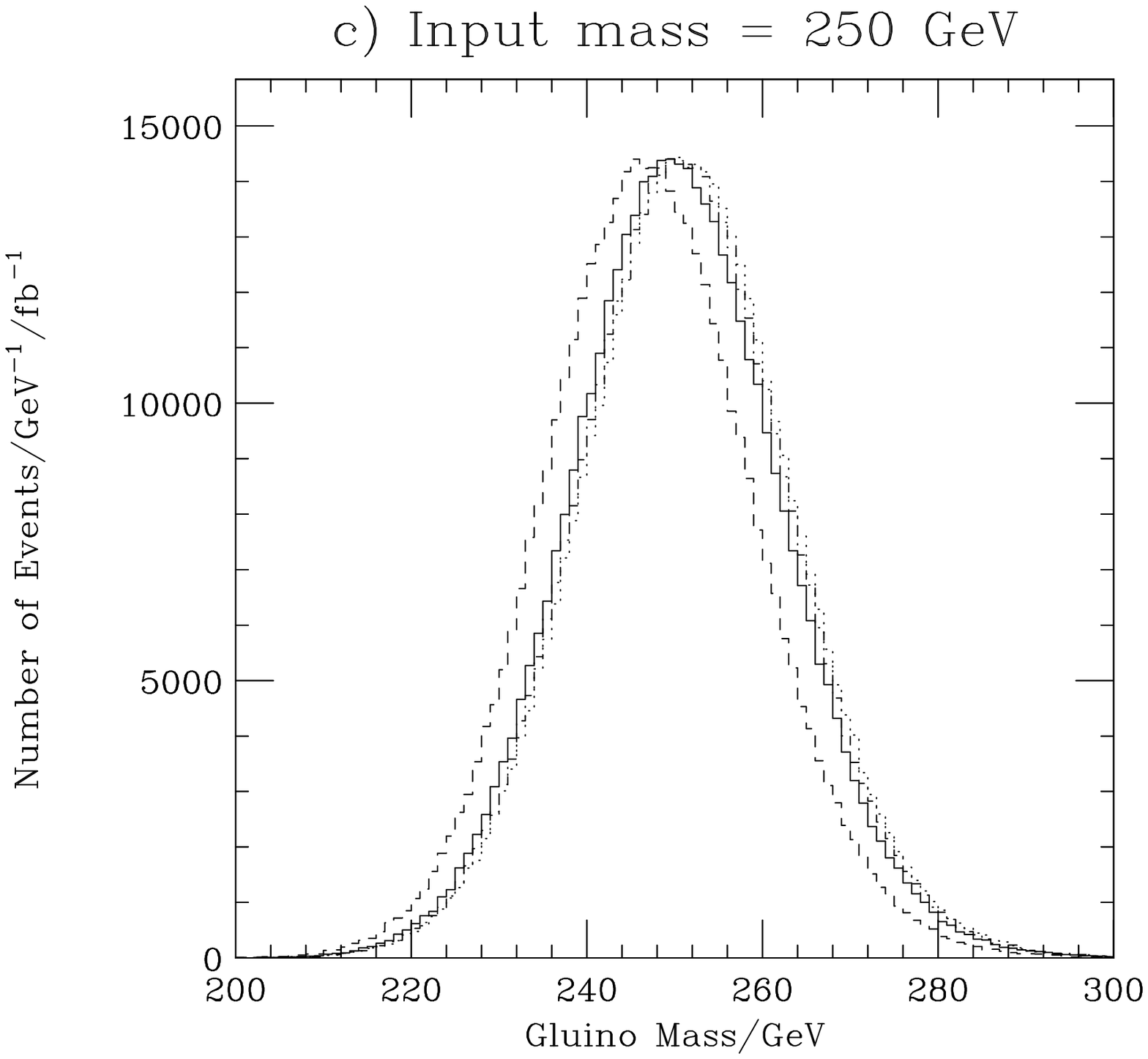}\hfill
\includegraphics[angle=90,width=0.45\textwidth]{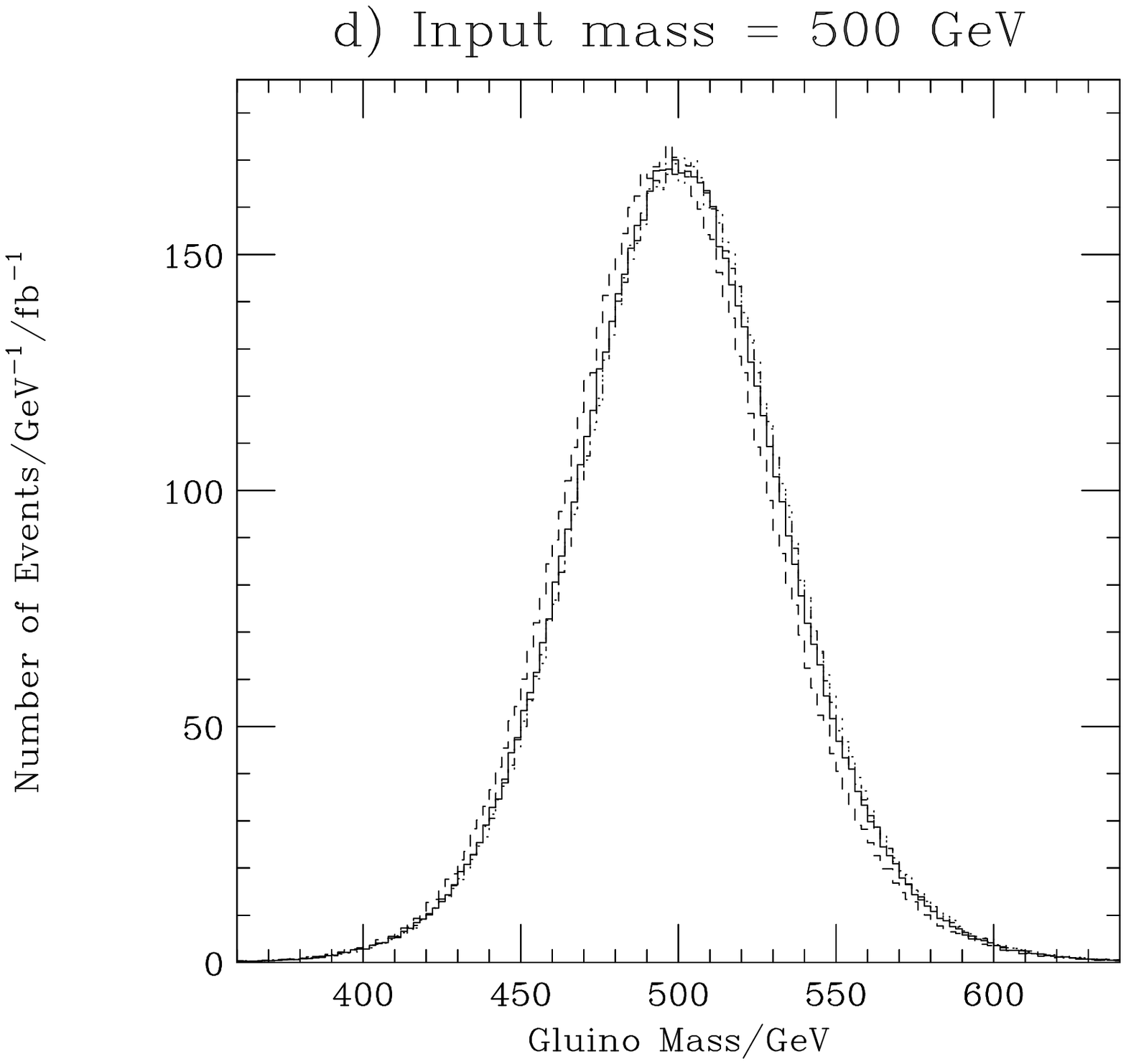}\\
\vspace*{-5mm}
\caption{\label{fig:mass1}
         Effect of the modelling of the interaction with the detector on the
         reconstructed $R$-hadron mass for different gluino masses.
         We show curves for $\lambda_{R}/2$ with the pomeron cross section form (solid),
         $\lambda_{R}/2$ with the cut-off form (dashed), 
         $\lambda_{R}$ with the pomeron form (dot-dashed)
         and $2\lambda_{R}$ with the pomeron form (dotted). The probability for
         producing the $R_g^0$ is set to zero. We simulate one million events,
         which is
         less than one year of high-luminosity running for the all the masses shown.}
\end{figure}

  The percentages of the different species of $R$-hadron
  is shown in Table~\ref{tab:Rproduction}
  for two different gluino masses. We see that the $R_{q\bar{q}}$
  hadrons containing only
  light quarks are predominantly produced with a preference for up and down quarks 
  over strange quarks.\footnote{The difference in production rates between the charged
                                 and neutral $\rho$ mesons would be corrected by the
                                 inclusion of
                                 the $R_\omega$ meson. Given our simple modelling
                                 this would not affect the results.} 
  The production rates for $R$-hadrons containing a heavy bottom
  or charm quark is very low.
  The reason for this is that in our simulation these mesons can only be produced if a gluon 
  that is colour-connected to the gluino perturbatively branches into
  a heavy-quark pair. This might be
  an underestimate of the production rate for these states; in the same way, HERWIG tends
  to underestimate the production rates for bottom and
  charmonium, which are produced by the same mechanism, at LEP energies.
  However, given the very low
  rates it is unlikely that there will be enough events to detect a signal based on displaced
  vertices due to the decay of the companion heavy quark. The information that
  could be extracted from these decay signatures would of course be highly interesting.

  While all these assumptions are necessary to perform the simulations, they do not
  have a major effect on the signals we will consider: the phenomenology is mainly
  determined by the gluino mass and the probability $P_{R_g}$ of producing the $R^0_g$
  rather than an $R$-meson in the cluster decays.

\begin{figure}[t]
\begin{center}
\includegraphics[angle=90,width=0.45\textwidth]{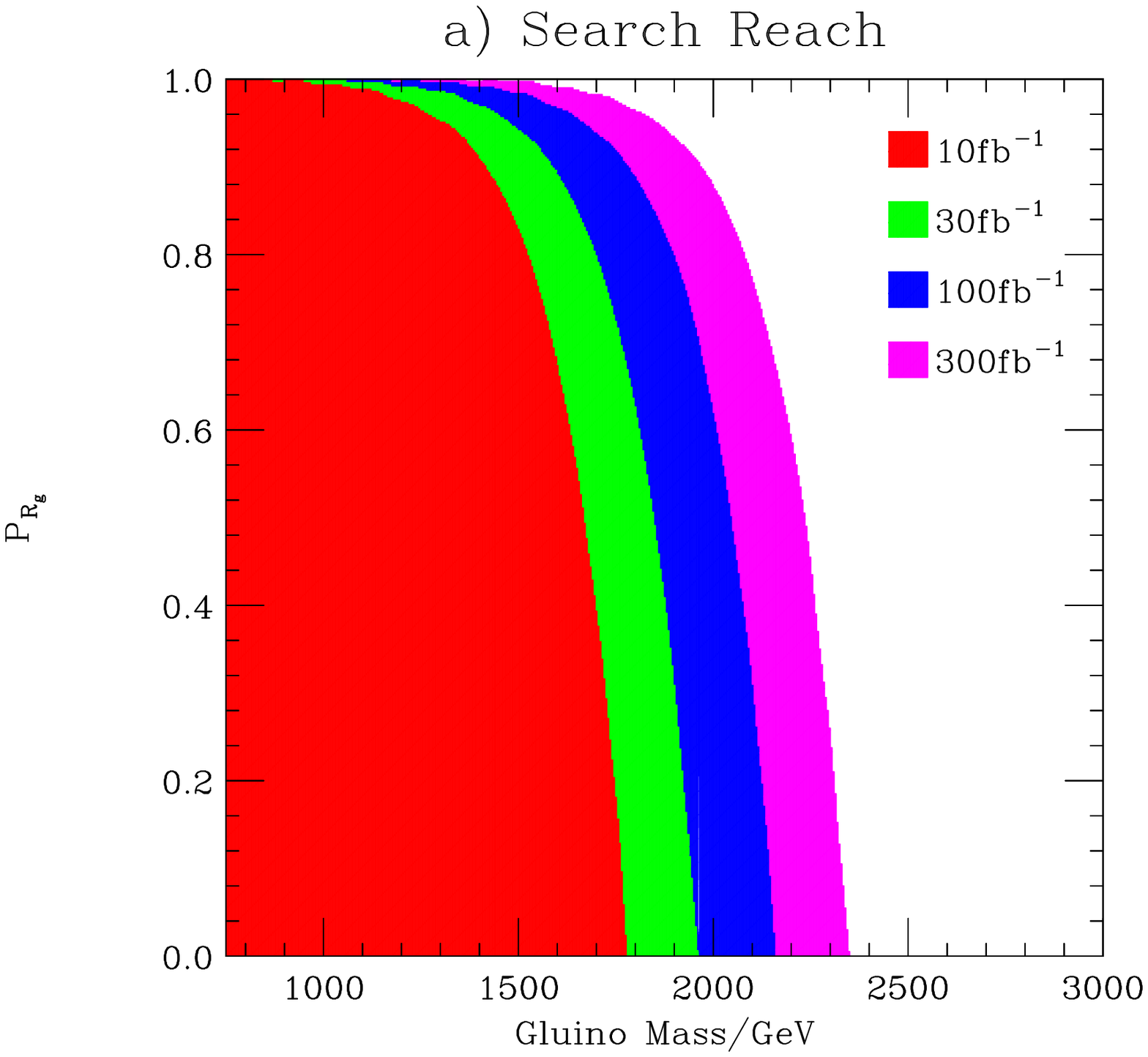}
\includegraphics[angle=90,width=0.45\textwidth]{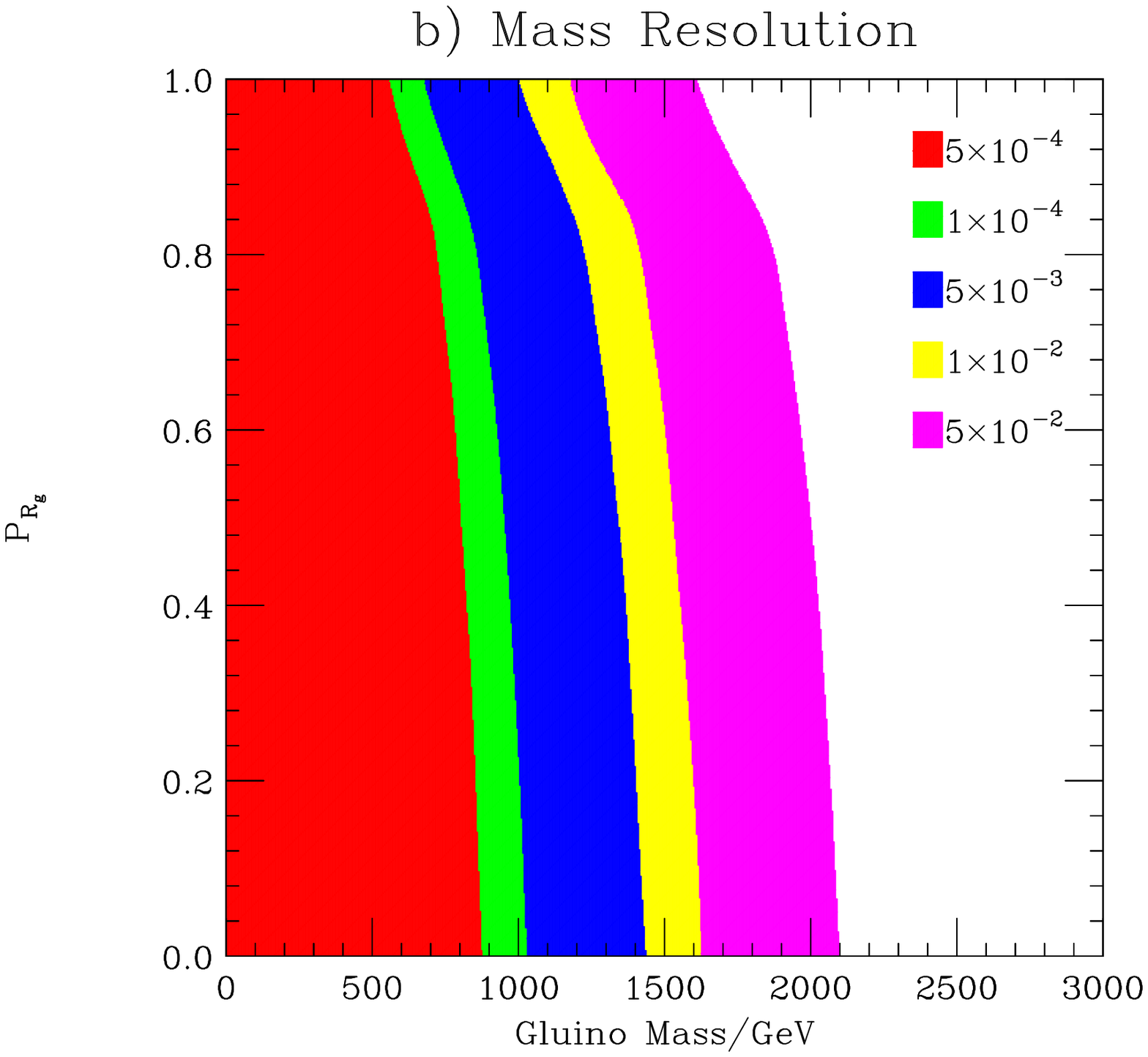}
\end{center}
\caption{\label{fig:chargedreach}
         Discovery reach (a) and mass resolution (b) for charged $R$-hadrons.
         We require the observation
        of ten charged $R$-hadrons for 
        four different integrated luminosities. We show the mass resolution
        $\Delta M/M$ for $100\fb^{-1}$.}
\end{figure} 

\begin{table}[t]
\begin{small}
\begin{center}
\begin{tabular}{|c|c|c|c|c|c|c|c|c|c|c|c|}
\hline
Variable & \multicolumn{10}{c|}{Allowed values}\\
\hline
$\not\!\!E_T\,[\GeV]$   & 100 & 150 & 200 & 300 & 400 & 600 & 800 & 1000 & 1500 & 2000\\
\hline
$P_{T_{j_1}}\,[\GeV]$   & 100 & 150 & 200 & 300 & 400 & 600 & 800 & 1000 & 1500 & 2000\\
\hline
$P_{T_{j_2}}\,[\GeV]$   &  50 & 100 & 150 & 200 & 300 & 400 & 500 & 600 & 800 & 1000\\
\hline
$\sum P_{T_j}\,[\GeV]$   & 100 & 200 & 300 & 400 & 500 & 600 & 800 & 1000 & 1500 & 2000\\
\hline
$N_{\rm jet}$           & 2 & 3 & 4 & 5 & 6 & 7 & 8 & 9 & 10 & 11\\
\hline
$S_T$                   & 0 & 0.1 & 0.2 & 0.3 & 0.4 & 0.5 & 0.6 & 0.7 & 0.8 & 0.9\\
\hline
$\Delta\phi_{j_1}$      & 0 & 0.3 & 0.6 & 0.9 & 1.2 & 1.5 & 1.8 & 2.1 & 2.4 & 2.7\\
\hline
\end{tabular}
\end{center}
\end{small}
\caption{\label{tab:etmisscuts} Allowed values for each of the cuts
         described in the text. In each case the variable is required
         to be larger than the value quoted.}
\end{table} 

\begin{figure}[t]
\includegraphics[angle=90,width=.45\textwidth]{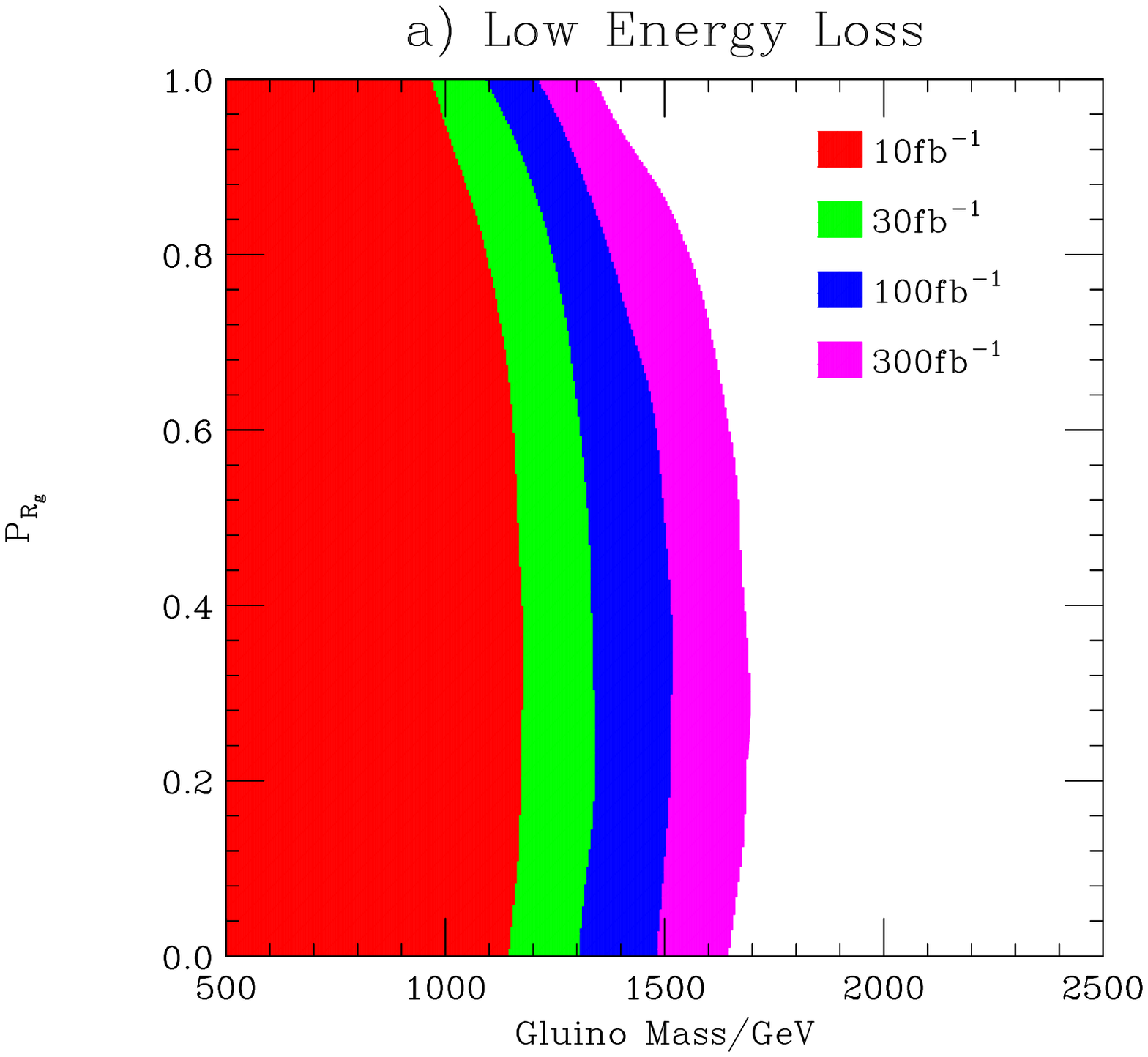}\hfill
\includegraphics[angle=90,width=.45\textwidth]{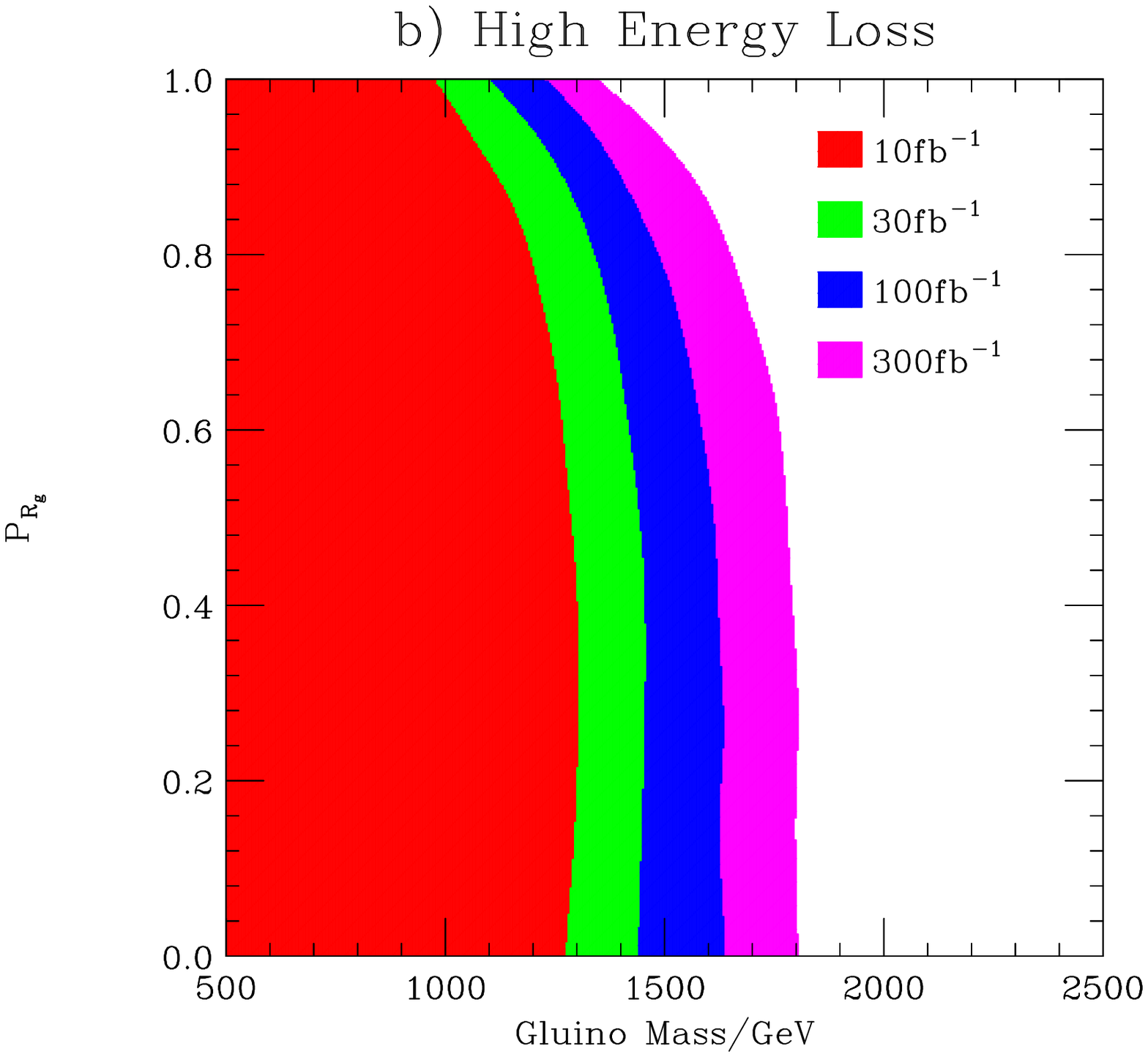}\\
\caption{\label{fig:etmissreach}
         Discovery reach for the missing transverse energy plus jets signal.
         (a) shows the discovery potential using twice the interaction length
         and the triple-pomeron form of the cross section, which leads to the
         smallest energy loss by the $R$-hadron. (b) shows the search reach using
         half the $R$-hadron interaction length and the cut-off form of the cross
         section, which leads to the largest energy loss by the $R$-hadron.}
\end{figure}

  In addition to the hadronization, we need to consider the interactions of the
  $R$-hadrons in the detector. For the interactions of the other particles 
  we use the AcerDet fast simulation~\cite{Richter-Was:2002ch}. 
  The interaction of the gluino is modelled in the same way as in Ref.~\cite{Baer:1998pg}.
  The energy and angular dependence of the $R$-hadron nucleon cross section
  are modelled using either the cut-off form with a cut-off value of $1\GeV$ or 
  the triple-pomeron form considered in Section~IIIA
  of Ref.~\cite{Baer:1998pg}. Rather than the approach taken in Ref.~\cite{Baer:1998pg},
  which  
  uses the average energy loss in these collisions combined with the depth of 
  the detector in terms of radiation lengths, we propagated the $R$-hadrons through
  the detector. We generate the distance to the next interaction according to the
  exponential distribution using the interaction length. The differential
  cross section is used to calculate the energy loss and change in direction of the 
  $R$-hadron due to the collision. This gives us fluctuations in the energy loss on
  an event-by-event basis. For the basic properties of the detector we use
  the parameters given in Table~\ref{tab:detector}, which are based on the ATLAS
  detector~\cite{atlastdrI}. The energy loss in the electromagnetic and hadronic 
  calorimeters is added to the cells of the calorimeter of the fast detector
  simulation. For charged $R$-hadrons, which will be detected in the muon chambers,
  we assume that the momentum measurement will be dominated by the momentum when the
  $R$-hadron reaches the muon chamber. The time of arrival of this meson at 
  the muon chamber is smeared with a Gaussian of width $0.7\,\rm{ns}$~\cite{Allanach:2001sd}
  and the momentum as described in Ref.~\cite{Allanach:2001sd}.\medskip

  Our signatures depend on the energy lost by the $R$-hadron
  in the detector. The simple cut-off ansatz for
  the cross section gives too much energy loss for incident pions,
  whereas the triple-pomeron form gives a good approximation for pion energies lower than
  $100\GeV$ and too little energy loss at higher energies~\cite{Baer:1998pg}. We investigate the
  uncertainty in the modelling of the interaction of the $R$-hadron with the
  detector in two ways. The simplest approach is to use the two different models of the
  cross section. The second is to vary also the interaction length $\lambda_R$ of the 
  $R$-hadron in the detector between half and twice the value used in Ref.~\cite{Baer:1998pg}:
  $\lambda_R \sim 16/9\lambda_\pi$. The effects of these variations are
  shown in Fig.~\ref{fig:energyloss}. We find that the size of the variation
  decreases as the $R$-hadron mass increases and the effect becomes negligible for the 
  masses we are interested in.\bigskip

\subsubsection{\boldmath Charged $R$-hadron searches}

  We consider two main strategies for the analysis. The first closely follows the analysis in
  Ref.~\cite{Allanach:2001sd} and requires the presence of a charged
  $R$-hadron, which is reconstructed as a muon. The transverse momentum of
  the hadron has to be larger than $50\GeV$ (which is sufficient 
  to trigger the event) and the time delay with
  respect to an ultra-relativistic particle $\Delta t$ has to satisfy $10\;{\rm ns}<\Delta t<50\;{\rm ns}$.
  An efficiency of 85\% is applied for the probability of reconstructing a muon~\cite{Allanach:2001sd}. 
  
  The mass of the $R$-hadron can then be reconstructed using
\begin{equation}
m^2=\frac{p_D\Delta t}{x}\left(2p+\frac{p_D\Delta t}{x}\right),
\end{equation}
  where $p$, $p_D$ and $x$ are the momentum, the transverse momentum and radius of the muon detectors 
  or the momentum along the beam direction and half-length of the muon detector,
  depending on whether the $R$-hadron hits the barrel or end-cap detectors.

  It is important to check what the effect of the modelling of the $R$-hadron interaction 
  with the calorimeter on the mass determination is. The
  reconstructed mass is shown for different gluino masses and
  choices of the interaction with the calorimeter in Fig.~\ref{fig:mass1}.
  As we expect from Fig.~\ref{fig:energyloss} the effects of the different
  choices of the $R$-hadron interaction length and cross section are more 
  apparent at low gluino masses. For all masses, halving the interaction
  length and using the cut-off form of the cross section leads to more energy loss
  by the $R$-hadron and hence a lower peak value for the mass and fewer
  events passing the cuts. For a gluino mass of $50\GeV$, the shift in the
  average mass is $2.3\GeV$, for a mass of $500\GeV$ the shift is $5\GeV$, and
  for a mass of $2\,\TeV$ is $6.7\GeV$. In a more realistic study this
  shift could be corrected for by 
  including the energy deposited in the calorimeter when measuring the $R$-hadron mass.

  The cuts we apply should eliminate the Standard Model
  background~\cite{Allanach:2001sd}. In order to calculate the
  discovery reach for charged $R$-hadrons we require the observation of 
  ten $R$-hadrons.
  The results shown in Fig.~\ref{fig:energyloss} are using half the default $R$-hadron interaction length and
  cut-off form of the $R$-hadron interaction cross section. This is the 
  model that gives the highest energy loss. However, the results are not 
  particularly sensitive to this choice and the choice of parameters with the lowest
  energy loss we consider gives only marginally better results.
  The reach for this signal is shown in Fig.~\ref{fig:chargedreach}
  in the $m_{\tilde g}$--$P_{R_g}$ plane. We see that the discovery reach extends to
  over $1.5\,\TeV$ for one year's running at low luminosity and to over $2\,\TeV$ for
  one year at high luminosity, apart from a region with low probability of producing
  a mesonic $R$-hadron.

  The resolution of the reconstructed $R$-hadron mass is shown in 
  Fig.~\ref{fig:chargedreach} for an integrated luminosity of $100\fb^{-1}$.
  For masses of less than $500\GeV$ the mass can be measured with a 
  precision of better than $0.05\%$. For these masses, this precision is better than
  the shift in the measured mass due to the energy loss of the charged $R$-hadron in
  the calorimeter. 
  For higher masses the resolution decreases, but it is still better than $1\%$ for masses
  up to $1.5\,\TeV$. This is an effect of the mass shift due to 
  the energy loss in the calorimeter.\bigskip

 \subsubsection{\boldmath Neutral $R$-hadron searches}

  A second signal that does not depend on the production of charged $R$-hadrons is
  the classic jets plus missing transverse energy signature.
  Neutral $R$-hadrons will of course always be produced, even 
  if no $R^0_g$ hadrons are created, because the $R_\rho^0$ will be produced
  with the same probability as the charged $R$-hadrons, see Table~\ref{tab:Rproduction}.
  For neutral $R$ hadrons only there will be no production
  of charged leptons in association with the gluino signal, apart from the decays
  of heavy hadrons. On the other hand, there will be fake muons from the charged $R$-hadrons.
  In analysing the missing transverse energy signal we therefore
  require that there
  be no leptons in the signal, so as to reduce the background from Standard
  Model $W$ and $Z$ production. When applying this cut we assume that charged 
  $R$-hadrons that pass the same isolation cut as muons will be reconstructed as muons.
  This is a conservative assumption, because some of them will not be reconstructed, because of
  the large time delay.

  Our approach for neutral $R$-hadrons is close to that in Ref.~\cite{Barr:2002ex}. The 
  Standard Model QCD, top quark, $W$+jets and $Z$+jets signals are simulated using 
  HERWIG6.5~\cite{Corcella:2000bw} in logarithmic transverse momentum
  bins in order to increase the number of events simulated at high-$p_T$, which 
  are most likely to contribute to the background. A number of variables are 
  used to distinguish between the signal and background events:\\\smallskip
\noindent
 (1) the missing transverse energy $\not\!\!E_T$;\\
 (2) the transverse momentum of the hardest jet $P_{T_{j_1}}$;\\
 (3) the transverse momentum of the second hardest jet $P_{T_{j_2}}$;\\
 (4) the scalar sum of the transverse momentum of the jets in the event $\sum P_{T_j}$;\\
 (5) the number of jets $N_{\rm jet}$;\\
 (6) the transverse sphericity of the event $S_T$;
      see \eg Ref.~\cite{atlastdrII} for the definition of $S_T$;\\
 (7) the difference in azimuthal angle between the direction of the hardest jet and $\not\!\!E_T$.\smallskip

  We test different sets of cuts on these variables to
  maximize the statistical significance of the signal on a point-by-point basis.
  The values of the cuts are given in Table~\ref{tab:etmisscuts}. The 
  lowest value of the cuts on the missing transverse momentum and on the transverse momentum of the 
  first jet are sufficient for the event to be triggered~\cite{Barr:2002ex}.
  We define the significance as $S/\sqrt{S+B}$,
  to minimize the effect of statistical fluctuations in the signal and 
  background samples and require a $5\sigma$ significance.\smallskip

  The discovery potential for this signal is shown in Fig.~\ref{fig:etmissreach} for 
  the largest and smallest energy loss by the $R$-hadron considered in our modelling of
  the $R$-hadron interaction with the detector. 
  For both choices of this interaction, the discovery
  potential is smallest for high probabilities of producing the $R_g^0$, and it increases with
  the probability of producing an $R$-meson. This is because 
  there can be significant missing transverse energy when 
  a neutral $R$-hadron is produced together with a charged one. 
  If the charged $R$-hadron is considered to be a jet with a non-isolated muon,
  this gives a jet and significant missing transverse 
  energy.
  This also explains why the model of the interaction with less energy loss gives a 
  lower signal: if the charged $R$-hadrons deposit less energy in the calorimeter, they
  are more likely to be considered as leptons and not included in the analysis, which 
  reduces the signal. 

  Even using this model of low interaction with the detector, gluinos
  with masses up to $1.1\,\TeV$ can be discovered with
  $100\,\rm{fb}^{-1}$ integrated luminosity, and gluinos with masses
  up to $1.3\,\TeV$ can be observed with an integrated luminosity of
  $300\,\rm{fb}^{-1}$.\footnote{Recently, the interactions of the
  $R$-hadrons in the detector has been considered in more
  detail~\cite{Kraan:2004tz}. While the energy losses for the
  $R$-hadrons they find is generally within the broad range we
  consider they find a significant probablity of the conversion of
  mesonic into baryonic $R$-hadrons, which we have neglected and may
  reduce the discovery potential for charged $R$-hadrons.} The main 
  difference between the searches for charged and neutral $R$-hadrons 
  is that in the missing transverse energy search we will not be able
  to measure the gluino mass except through the total cross section.
  However, this might be possible for gluino masses of ${\cal O}(\TeV)$, 
  if we look for gluino--gluino bound states leading to a peak in 
  the two-jet invariant mass spectrum~\cite{Kuhn:1983sc}.
 
\begin{figure}[t]
\begin{center}
\includegraphics[width=0.60\textwidth]{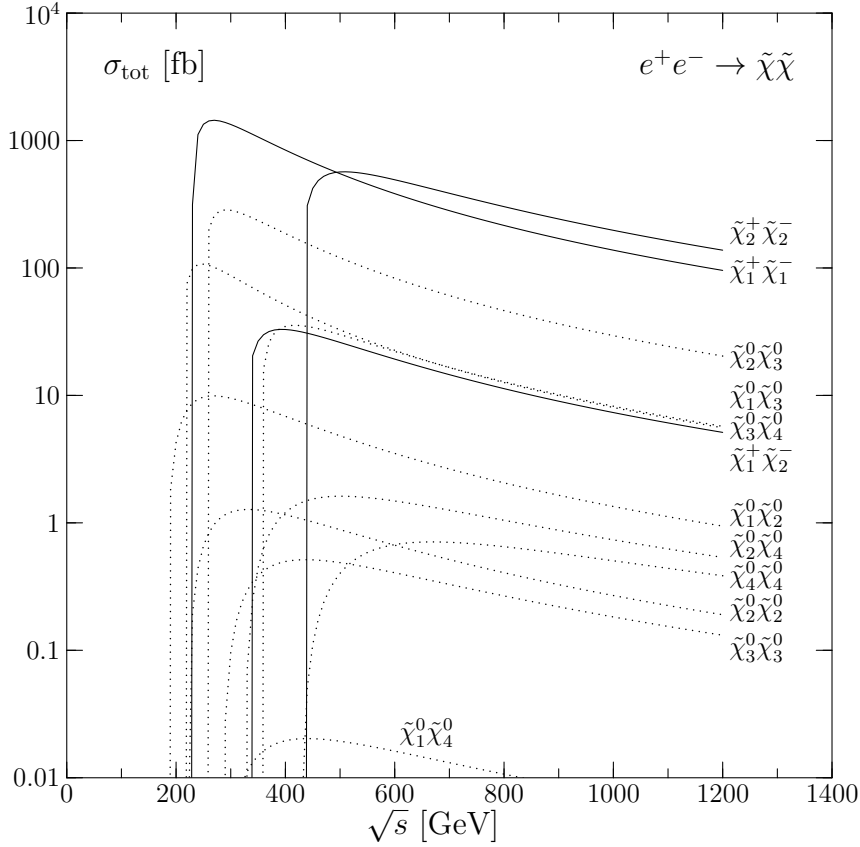}
\end{center}
\vspace{5mm}
\caption{\label{fig:prod-LC} Cross sections for chargino and
neutralino pair production in $e^+e^-$
collisions~\cite{Kilian:2001qz}, assuming the reference
point in eq.(\ref{eq:sps1}).}
\end{figure}

\section{Yukawa couplings from gaugino--Higgsino mixing}

If split supersymmetry should be realized in nature, the observation
of the gluino, charginos and neutralinos will only be the
first task. Once these states are discovered, we will have to show
that they constitute a weak-scale SUSY Lagrangian. At the LHC, the
immediate challenge will be the determination of the quantum numbers
of the new particles~\cite{Barr:2004ze}. Even if we take for
granted their fermionic nature, this does not establish them as SUSY
partners: the set of colour-octet, weak isosinglet, isotriplet, and a
pair of isodoublet fermions (as present in the MSSM) makes up a
minimal non-trivial extension of the Standard Model that is anomaly-free and
consistent with gauge coupling unification. A quantitative hint for
supersymmetry is given by the off-diagonal elements in the mass
matrices. They determine the mixing of gauginos and higgsinos into 
charginos and neutralinos as mass eigenstates. This mixing is possible 
because supersymmetry transformations maintain Standard Model 
quantum numbers. These off-diagonal entries in the mass matrix also 
constitute the neutralino and chargino Yukawa couplings. 
In SpS, these off-diagonal
entries follow, up to renormalization group effects, the predicted
MSSM pattern.\medskip

Without any mixing, the only production channels (in the absence of 
scalars) in $\bar q q$ or $e^+
e^-$ annihilation are $\tilde\chi^+_1\tilde\chi^-_1$,
$\tilde\chi^+_2\tilde\chi^-_2$, $\tilde\chi^\pm_1\tilde\chi^0_2$,
$\tilde\chi^\pm_2\tilde\chi^0_{3,4}$, and
$\tilde\chi^0_3\tilde\chi^0_4$. 
Moreover, all these produced particles would be stable. 
Any other production or decay channel requires either a finite 
coupling to $s$ channel gauge bosons through mixing or the 
presence of scalars. 
The observation of additional production channels and
the measurement of decay branching ratios is therefore an indirect
probe of the neutralino and chargino Yukawa couplings. 
The usual analysis of gauge couplings and the corresponding 
gaugino-sfermion-fermion couplings will fail in SpS scenarios, because
the squarks are much heavier than the gauginos~\cite{Cheng:1997vy}.
To measure the
neutralino and chargino mixing matrices, a precise mass measurement is
sufficient. Without gaugino--Higgsino mixing the mass matrices are
determined by the MSSM parameters $M_1, M_2$ and $\mu$. The
gaugino--Higgsino mixing adds terms of the order of $M_Z$ and
introduces the additional parameter $\tan\beta$, leading to four MSSM
parameters altogether. As shown
before~\cite{Feng:1995zd}, these parameters can be
extracted from the six neutralino and chargino masses by using a simple
fit, properly including experimental errors~\cite{Barger:1999tn}.

Figure~\ref{fig:prod-LC} displays the cross sections for chargino and
neutralino pair production in $e^+e^-$ collisions for the point of
eq.(\ref{eq:sps1}) as a function of the collider energy. With one
exception, all channels have cross sections larger than $0.1\fb$ and
the threshold value for $\tilde\chi^+_1\tilde\chi^-_1$ production is
as large as $1\;\pb$. A linear collider with moderate energy and high
luminosity would be optimal to probe all these processes, and some kind
of fit is the proper method to extract the weak-scale Lagrangean
parameters. We emphasize that these cross
sections~\cite{Oller:2004br} as well as the
masses~\cite{Fritzsche:2002bi} are known to NLO. However, because we
are mainly interested in the error on the extracted underlying
parameters and less interested in their central values, we limit our
fit to leading order observables.
\medskip

Previous studies of the chargino and neutralino systems
concentrated on the extraction of the mass parameters $M_1,M_2,\mu$,
while the off-diagonal elements were fixed or at least related to each
other by the MSSM relations. In SpS, the four off-diagonal entries in
the mass matrices are independent observables. As defined in
eq.(\ref{eq:sps1_kappa}) we parametrize the couplings $\tilde
g_{u,d}^{(\prime)}$, introducing an additional factor
$(1+\kappa_{u,d}^{(\prime)})$ with respect to the MSSM
values~\cite{Giudice:2004tc}. While these $\kappa$ parameters vanish
to leading order in the complete weak-scale MSSM, the SpS
renormalization flow between the matching scale $\sms$ and the
electroweak scale induces non-zero values of order
$\kappa^{(\prime)}_i = -0.2\ldots 0.2$.  If we are
able to detect deviations of this size at a collider, we can both
establish the supersymmetric nature of the model and verify the
matching condition to the MSSM at $\sms$.\medskip

\begin{figure}[t]
\begin{center}
\includegraphics[width=16.0cm]{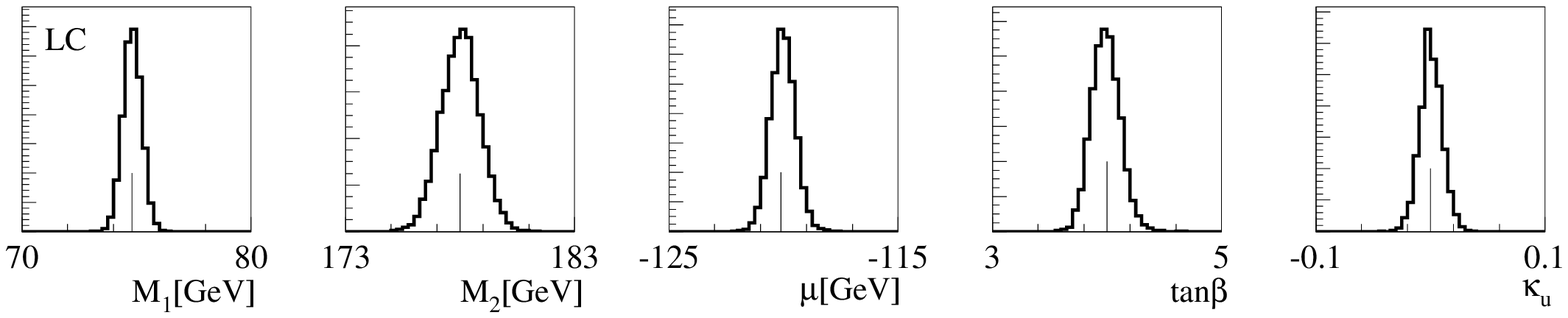} \\
\includegraphics[width=16.0cm]{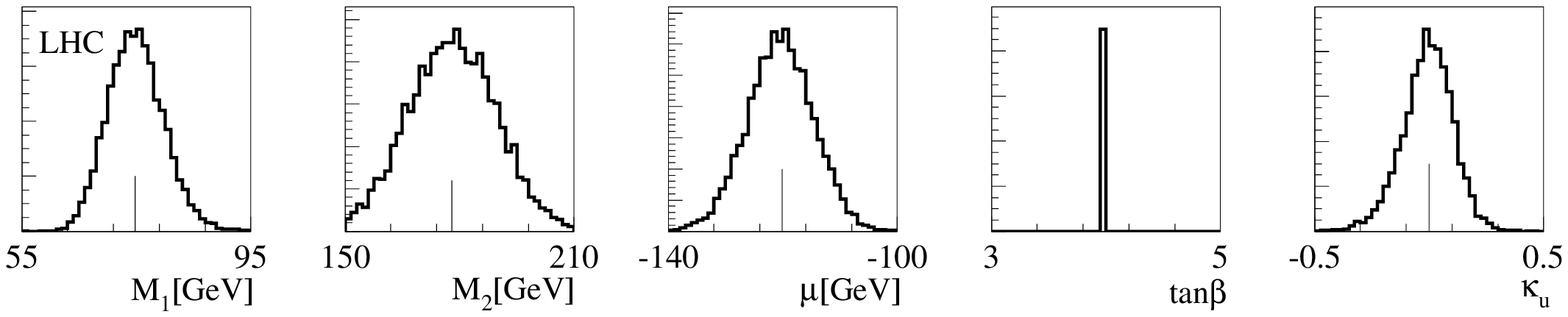}
\end{center}
\vspace{-8mm}
\caption{\label{fig:fit_1} Fit to 10000 sets of mass and cross section
pseudo-measurements at a future linear collider (upper) and at the LHC
(lower). The fitted parameters include only $\kappa_u$ with a central
value zero. At the LHC $\tan\beta=4$ is fixed.}
\end{figure}

As mentioned above, the neutralino and chargino mixing matrices can be
measured at a future linear collider, in continuum production as well
as through a threshold scan~\cite{Aguilar-Saavedra:2001rg}. The six
masses alone are sufficient to determine all the usual MSSM parameters
$M_1,M_2,\mu,\tan\beta$, plus one additional $\kappa$. Because the
predicted values $\kappa^{(\prime)}_i$ are small, the correct
treatment of the experimental accuracies is crucial. For our central
parameter point of eq.(\ref{eq:sps1}) we compute the masses and the cross
sections shown in Fig.~\ref{fig:prod-LC}, with the exception of the
$\tilde\chi^0_1 \tilde\chi^0_1$ channel. To all observables we assign
an experimental error, which in our simplified treatment is a relative
error of $0.5\%$ on all linear-collider mass
measurements~\cite{Aguilar-Saavedra:2001rg}, $5\%$ on
all LHC mass measurements~\cite{Bachacou:1999zb}, and
the statistical uncertainty on the number of events at a linear
collider corresponding to $100 \fb^{-1}$ of data at a $1 \TeV$
collider after all efficiencies\footnote{Disentangling the various
channels is a non-trivial task, but in the final fit it is always
possible to replace the total cross sections by any other
measurement. We leave this complication to a more detailed study of
the experimental uncertainties~\cite{Aguilar-Saavedra:2001rg} and the
proper correlated fit including statistical, systematical and
theoretical errors~\cite{Lafaye:2004cn}.}. The assumption that the
LHC might be able to see all six gauginos and Higgsinos and measure 
all their masses is very optimistic, so we will only use it to derive 
the maximum sensitivity the LHC could achieve.

Moreover, the precision
of the theoretical cross section prediction~\cite{Beenakker:1996ch}
and the precision of the measurement of cross sections and branching
ratios at the LHC is likely to be insufficient to allow the extraction of small mixing
parameters. Around the central parameter point we randomly generate
10000 sets of pseudo-measurements, using a Gaussian smearing. Out of
each of these sets we extract the MSSM parameters. In principle, we
could simply invert the relation between the masses and the Lagrangian
parameters analytically. However, after smearing, this inversion will not
have a unique and well-defined solution; therefore we use a fit to
solve the overconstrained system. The distribution of the 10000 fitted
values should return the right central value and the correctly
propagated experimental error on the parameter determination. If
necessary, we apply another maximum $\chi^2$ cut on the 10000 fits, to
get rid of secondary minima. The distributions of the measurements are
not necessarily Gaussian, and there might be non-trivial correlations
between different measurements. At the end, the crucial questions are:
(i) is the error on the parameter measurements sufficient to claim
agreement with the MSSM prediction, and (ii) Is the measurement good
enough to probe the renormalization group effects of the heavy scalars
in SpS?\smallskip

\begin{table}[t]
\begin{center}
\begin{tabular}{|l||c|c|c||r|r|r|r|}
\hline
          & Fit $\tan\beta$   & $m_i$    & $\sigma_{ij}$ & $\Delta \kappa_u$   & $\Delta \kappa_d$ & $\Delta \kappa'_u$  & $\Delta \kappa'_d$\\

\hline \hline
Tesla     &                   & $\bullet$ & $\bullet$      & $0.9 \times 10^{-2}$ & $3 \times 10^{-2}$ & $1.3 \times 10^{-2}$ & $4  \times 10^{-2}$ \\
Tesla     & $\bullet$          & $\bullet$ & $\bullet$      & $1.2 \times 10^{-2}$ & $5 \times 10^{-2}$ & $2   \times 10^{-2}$ & $5  \times 10^{-2}$ \\
Tesla     &                   & $\bullet$ &               & $1.1 \times 10^{-2}$ & $5 \times 10^{-2}$ & $3   \times 10^{-2}$ & $8  \times 10^{-2}$ \\
Tesla     & $\bullet$          & $\bullet$ &               & $1.2 \times 10^{-2}$ & $11\times 10^{-2}$ & $4   \times 10^{-2}$ & $8  \times 10^{-2}$ \\
LHC       &                   & $\bullet$ &               & $2.2 \times 10^{-1}$ & $6 \times 10^{-1}$ & $2.7 \times 10^{-1}$ & $8  \times 10^{-1}$ \\
\hline \hline
Tesla     &                   & $\bullet$ & $\bullet$      & $1.4 \times 10^{-2}$ & $5 \times 10^{-2}$ & $3   \times 10^{-2}$ & $10 \times 10^{-2}$ \\
Tesla$^*$ & $\bullet$          & $\bullet$ & $\bullet$      & $1.7 \times 10^{-2}$ & $9 \times 10^{-2}$ & $4   \times 10^{-2}$ & $13 \times 10^{-2}$ \\
Tesla     & fix $\tan\beta=3$ & $\bullet$ & $\bullet$      & $1.6 \times 10^{-2}$ & $4 \times 10^{-2}$ & $4   \times 10^{-2}$ & $9  \times 10^{-2}$ \\
Tesla$^*$ & $\kappa_i \ne 0$  & $\bullet$ & $\bullet$      & $1.4 \times 10^{-2}$ & $5 \times 10^{-2}$ & $4   \times 10^{-2}$ & $11 \times 10^{-2}$ \\
\hline
\end{tabular}
\end{center}
\caption{\label{tab:fit} Error on the determination of $\kappa_i$
from measured masses and possibly production cross sections. For the
first five lines, all but one $\kappa$ are fixed to zero, the fitted
$\kappa$ has the central value zero. In the last four lines, all four
$\kappa_i$ are fitted simultaneously. The very last line assumes the
predicted central values of $\kappa_i$ in our SpS parameter point. The
error on the mass measurements is $0.5\%$ for Tesla and $5\%$ for
the LHC. The sets of measurements marked by $^*$ include a maximum
$\chi^2$ cut to get rid of secondary minima.}
\end{table}

As a first test of our approach we set all four non-MSSM contributions
to zero ($\kappa^{(\prime)}_i=0$) and add one of the four anomalous Yukawa couplings to the
set of fitted parameters, keeping the other three fixed during the
fit. In Fig.~\ref{fig:fit_1} we show the result for a combined fit of
$M_1,M_2,\mu,\kappa_u$ and possibly $\tan\beta$. At a linear collider
we can extract the mass parameters at the percentage level and the
best measured anomalous coupling, $\kappa_u$, to typically $0.01$. In
Table~\ref{tab:fit} we see that the error on the determination of all
four $\kappa$ values at a linear collider is a few
per cent. Generically, the error on $\kappa^{(\prime)}_d$ is larger
than the error on $\kappa^{(\prime)}_u$, because $\kappa^{(\prime)}_d$
is accompanied by $\cos \beta$ while $\kappa^{(\prime)}_u$ enters with
an additional factor $\sin \beta$. We checked that for large $\tan\beta$ values,
\eg  $\tan\beta= 30$, only $\kappa^{(\prime)}_u$ can be extracted
with a reasonable error. If we fix all but one $\kappa$ to their zero
MSSM prediction, the remaining off-diagonal entries in the mass
matrices are determined by $\tan\beta$. While we might hope to extract
$\tan\beta$ from the Higgs sector, we also test the prospects of
determining it in our fit. In Table~\ref{tab:fit} we see that errors
only slightly degrade when we include $\tan\beta$ in the set of
parameters we fit to, and in Fig.~\ref{fig:fit_1} we see that the
determination of $\tan\beta$ indeed works very well.

Since we are limiting the number of unknowns to four or five
(depending on whether or not we fit $\tan\beta$), the six mass
measurements should be sufficient to extract one anomalous Yukawa coupling
parameter. Indeed, in Table~\ref{tab:fit} we see that the precision
on the $\kappa^{(\prime)}_i$ suffers only slightly when we limit our
set of measurements to the masses alone and assume $\tan\beta$ to be
known. This is an effect of the overwhelming precision of the mass
measurement through threshold scans,
our assumed error of $0.5\%$ is even conservative. Adding
$\tan\beta$ to the fit shows, however, that with five parameters and
six measurements our analyses are starting to lose sensitivity. When we try to
extract the Lagrangian parameters from a set of mass measurements at
the LHC, the errors on the mass parameters $M_1,M_2,\mu$ inflate to the
$10\% \ldots 20\%$ level, as shown in Fig.~\ref{fig:fit_1}. While we
might still be able test if the $\kappa^{(\prime)}_u$ follow the weak-scale MSSM
prediction, the experimental precision is clearly insufficient to
test the SpS renormalization group effects. Moreover, it is not clear
if all neutralino and chargino masses could be extracted at the LHC,
because all current search strategies rely on squark and gluino
cascade decays~\cite{Bachacou:1999zb}. Last but not
least, we do not know if we will be able to measure $\tan\beta$ in the
Higgs sector, and including $\tan\beta$ in the LHC fit will make the
extraction of $\kappa^{(\prime)}_u$ even less promising.\medskip

\begin{figure}[t]
\begin{center}
\includegraphics[width=16.0cm]{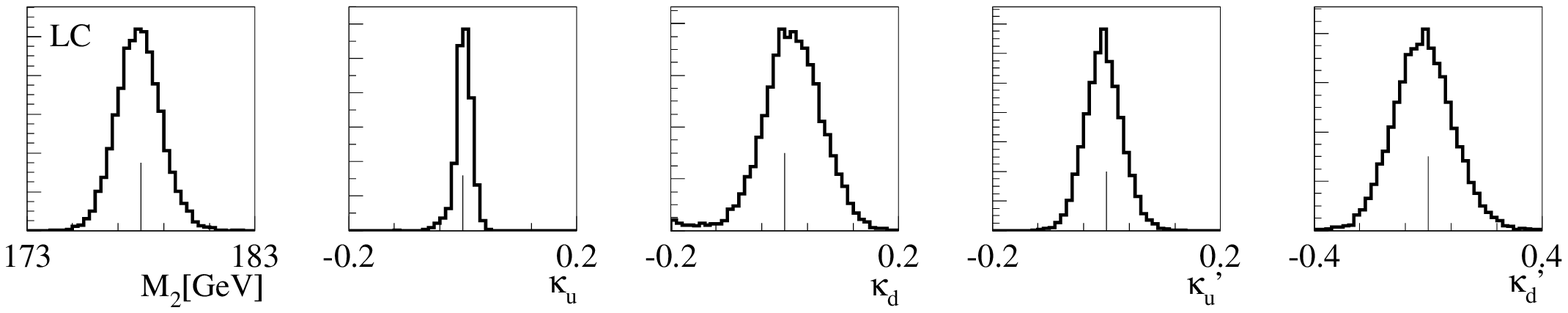} \\
\includegraphics[width=16.0cm]{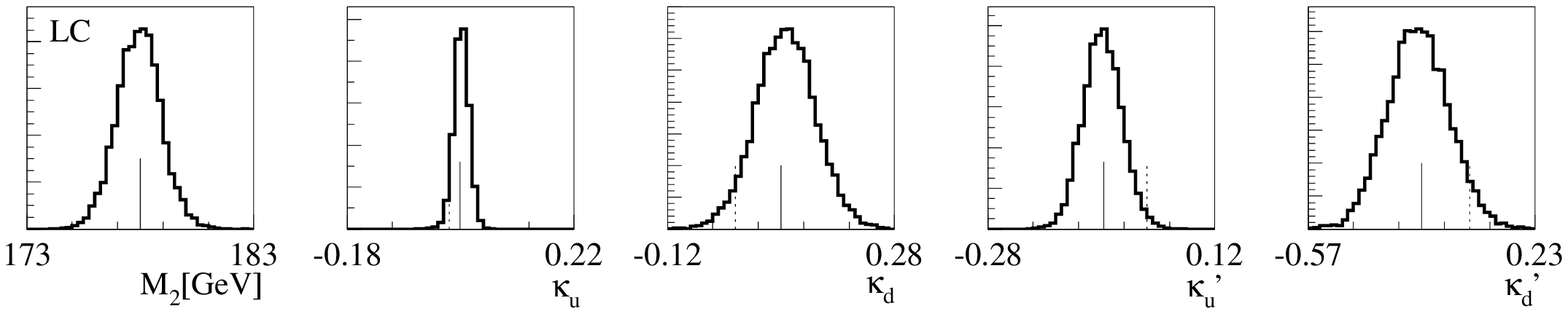}
\end{center}
\vspace{-8mm}
\caption{\label{fig:fit_2} Fit to 10000 sets of mass and cross section
pseudo-measurements at a future linear collider. All four
$\kappa^{(\prime)}_i$ are extracted simultaneously. The central values
are set to zero as in the MSSM (upper) and to the example SpS values
(lower). The MSSM zero prediction is indicated in the lower line of
histograms.}
\end{figure}

At the linear collider, adding the cross sections as independent
measurements allows us to fit all four $\kappa^{(\prime)}_i$
simultaneously. This is the proper treatment, unless we would have
reasons to believe that some of the $\kappa^{(\prime)}_i$ are predicted
to be too small to be measured. This means that $\tan\beta$ is no longer an
independent parameter: we can fix it in the fit, to reduce the
number of unknown Lagrangian parameters. The error on the
determination of all four $\kappa^{(\prime)}_i$ is shown in
Fig.~\ref{fig:fit_2}, including the error on $M_2$, to illustrate that
adding all four anomalous couplings to the fit has little impact on the
measurement of the dominant Lagrangian mass parameters. In
Table~\ref{tab:fit} the errors for the simultaneous $\kappa$
measurements are compared with the single-$\kappa$ fit. If we fix
$\tan\beta$ to the correct value, the error bands increase by a factor of
2.5 at the maximum, when we move to a combined extraction of all
$\kappa^{(\prime)}_i$. Adding $\tan\beta$ to the fit shows us to which
degree we are already limited by the number of useful measurements:
the quality of the measurements suffers considerably and we have to
avoid secondary minima. However, as we already pointed out,
$\tan\beta$ should be fixed if we limit ourselves to independent
parameters. The question to know, what happens if we fix it to a wrong
value. From eq.(\ref{eq:sps1_kappa}) we see that assigning a wrong
value to $\tan\beta$ should just move the central values of the
extracted $\kappa^{(\prime)}_i$, in our case away from zero. As an
example, for an assumed value $\tan\beta$ the four anomalous coupling measurements
are centred around $0.023,-0.23,0.023,-0.23$ instead of zero, in the
order of Table~\ref{tab:fit}. Again in Table~\ref{tab:fit} we see
that the effect on the errors is indeed negligible.\smallskip

The last step left is from the case $\kappa^{(\prime)}_i=0$ to the
values predicted by SpS, eq.(\ref{eq:sps1_kappa}). This is merely a cross-check,
because the error in the extraction of the anomalous couplings should not
significantly depend on their central values. Indeed, in
Fig.~\ref{fig:fit_2} and in Table~\ref{tab:fit} we see that the
central value has no visible effect on the errors, even though in our
case it makes the fit more vulnerable to secondary minima. As usual,
we get rid of the secondary minimum using a maximum-$\chi^2$ value for
the 10000 pseudo-measurements, reducing the number of entries in the
histogram by $17\%$. These results for the linear collider indeed
indicate that we could not only confirm that the Yukawa couplings and
the neutralino and chargino mixing follow the predicted MSSM pattern;
for the somewhat larger $\kappa'_i$ values we can even
distinguish the complete weak-scale MSSM from a SpS spectrum.

\section{Direct measurement of Yukawa couplings}

\begin{figure}[t]
\begin{center}
\includegraphics[width=10cm]{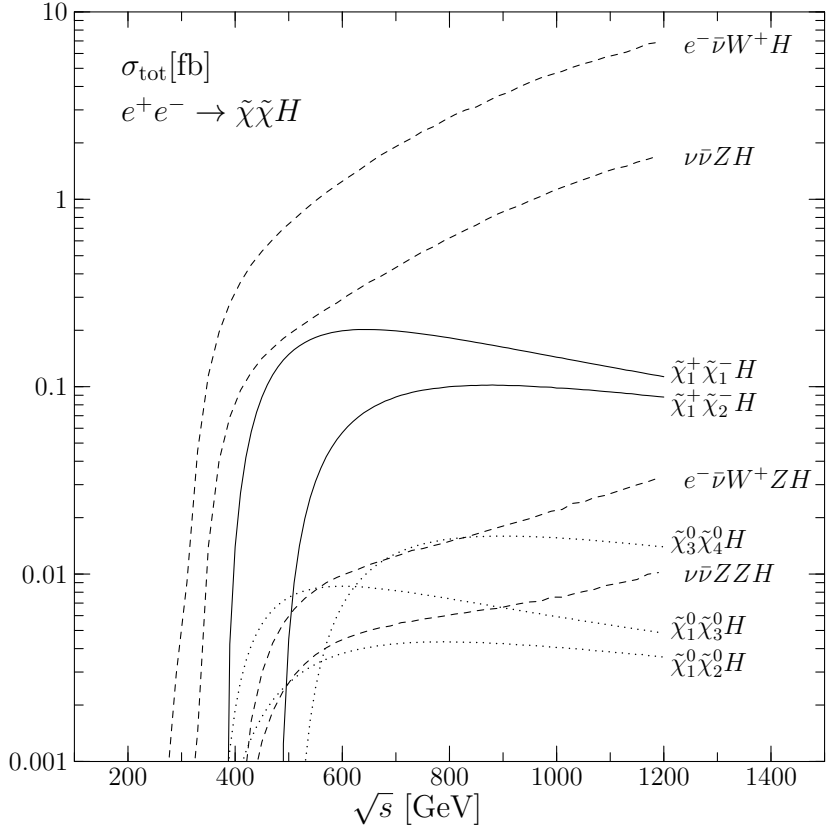}
\end{center}
\vspace{5mm}
\caption{\label{fig:cch-LC} Signal and background cross sections for
associated production of neutralinos and charginos with a Higgs boson
in $e^+e^-$ collisions~\cite{Kilian:2001qz}.}
\end{figure}

While the elements of the neutralino and chargino mixing matrices depend
on the Yukawa couplings in a complicated way, the cross sections for
chargino/neutralino pair production in association with a Higgs boson
are directly proportional to these parameters.  Therefore, the
observation of $H\tilde\chi\tilde\chi$ final
states~\cite{Kilian:2001qz,Ferrera:2004rf} could add to our knowledge
of the Yukawa couplings that can be gained in chargino/neutralino
pair production.\smallskip

If decays of the kind $\chi^\pm_2\to\chi^\pm_1H$ or $\chi^0_j\to\chi^0_iH$ proceed
with a significant rate, these branching fractions are determined by
the Yukawa couplings in conjunction with the mixing effects and should
be included in a global fit.  Unfortunately, for low values of the
mass parameters $M_1,M_2,\mu$ this is less likely to happen than in
the MSSM, since the Higgs is considerably heavier, so that some
channels are no longer kinematically accessible.  In particular, for
the reference point eq.(\ref{eq:sps1}), no such decay is possible.

However, in this situation there is still associated production of
charginos and neutralinos with a Higgs boson in the continuum, which
can in principle be observed at a high-luminosity $e^+e^-$
collider.  Figure~\ref{fig:cch-LC} displays the
cross sections as a function of the collider energy.  Like all
$s$-channel processes, the curves peak immediately above the
respective threshold energies.  The dominant Standard Model background
to these processes consists of Higgs production in association with
$Z$ and $W$ bosons and a neutrino pair, also shown in
Fig.~\ref{fig:cch-LC}.  The neutrino pair can either originate from a
$Z$ boson or from the continuum ($W$-fusion processes), where in the
latter case the total cross section increases with energy. Last but
not least, we have to take into account processes with a forward-going
electron in the final state, which may escape undetected.\medskip

Although the backgrounds are substantial, they do not affect all
signals simultaneously, and they can be reduced significantly by
kinematical cuts.  Assuming thus that processes with Higgs-strahlung
off a chargino or neutralino can be identified above the background,
they will depend on the $\kappa$ directly and through the neutralino
and chargino masses.  Apparently, the neutralino processes have a rate
too small to help in disentangling the parameters.  However, in
contrast to chargino pair production, $\tilde\chi^+\tilde\chi^-H$
associated production depends on $\kappa_u$ and is independent of
$\kappa'_u$, so its inclusion in a global fit reduces the correlation
between these two parameters. Because the dependence of the cross
sections on $\kappa'_d$ is fairly weak, this parameter will pose a
challenge for the direct extraction.

The precision that can be achieved from cross section measurements is
given by the statistical error on the cross section. Assuming
$1\ab^{-1}$ of integrated luminosity, we could collect a sample of at
maximum 200 (100) signal events, which gives us an error of $7\%
(10\%)$ on the cross section measurement and therefore an error of
$3.5\% (5\%)$ on the measurement of the Yukawa coupling. This number
could be competitive with our estimates for the indirect measurement
shown in Table~\ref{tab:fit}. However, even if we can extract the
masses involved using a threshold scan, the extraction of couplings
from cross sections is always plagued by theoretical uncertainties due
to higher orders and systematical experimental uncertainties. More
detailed studies are required to obtain a final verdict on the
errors~\cite{Lafaye:2004cn}.

\section{Conclusions}

Recently, models of split supersymmetry have been suggested. If we are
willing to accept a high degree of fine-tuning for the separation
between the weak scale and the Planck scale, decoupling of all
sfermions can solve problems which usual supersymmetry has in the flavour
sector, mediating proton decay or leading to large electric dipole
moments. In particular, gauge-coupling unification and the existence
of a dark matter candidate naturally survive the decoupling of the scalar partner
states.\smallskip

For collider experiments this means that only gauginos and Higgsinos
are light enough to be produced, because their masses can be protected
by a chiral symmetry. At the LHC we will observe a long-lived
gluino. Over almost the entire parameter space, we will be able to see
the resulting charged $R$-hadrons for gluino masses larger than
$2\TeV$ and determine the gluino mass to better than $1\%$. In the
region where the probability of producing a mesonic $R$-hadron is
small, the reach can be enhanced by the classic jets plus missing
energy channel. In the case of neutral $R$-hadrons this leaves us with a reach of between $1.3\TeV$ and
$1.8\TeV$ for the gluino mass, depending on the details of the
$R$-hadron spectroscopy. Because we cover neutral as well as charged 
$R$-hadrons our result is independent of the mass hierarchy of the
$R$-hadrons and the possible $R$-hadron decays into each other.
We emphasize that the interaction of the
gluino in the detector is currently being studied in more detail and we
expect improved estimates for the discovery reach as well as for the
mass measurement~\cite{Kraan:2004tz}.

Because cascade decays of squarks
and gluinos leading to subsequent neutralino and chargino signals will
not be available for split supersymmetry models, we give the direct
production cross sections for all possible channels. It will be a challenge to extract these
signals from the Standard Model backgrounds and separate them to gain
access to some of the model parameters.\smallskip

Obviously, a future high-luminosity linear collider will be perfectly
suited for this kind of precision measurements. Integrating out the
scalars at a high scale leads to renormalization group effects for the
neutralino and chargino Yukawa couplings and their mixing matrices. At
a linear collider we will be able to see all neutralinos and charginos
and measure their masses and cross sections, provided the collider energy is sufficient. From these measurements
we can extract the mixing matrix elements (with contributions from the anomalous Yukawa couplings) to better than $10\%$
accuracy. The estimate of the experimental errors is based on a similar 
parameter point studied in Ref.~\cite{Aguilar-Saavedra:2001rg}. For a 
final statement about the possible accuracy with which we can 
extract the anomalous Yukawa couplings one would have to combine a 
detector simulation with the proper treatment of the theoretical errors,
which is beyond the scope of this paper.
Through the associated production of charginos with a Higgs
we might also have direct access to these Yukawa couplings. If split
supersymmetry effects are large enough we will be able not only to
confirm that neutralinos and charginos are indeed the partners of
gauge bosons and Higgs bosons --- we will also gain insight into the
heavy decoupled spectrum.

\subsection*{Acknowledgements}

First, we would like to thank Andrea Romanino for his great help cross-checking 
our results with Ref.~\cite{Giudice:2004tc}. Moreover, we
would like to thank J\"urgen Reuter for providing the MSSM code for
the WHIZARD/O'Mega package and for reading the manuscript. We
would like to thank Aafke Kraan, Tim Jones, Peter Zerwas, Bryan
Webber and Andy Parker for fruitful discussions, which had a major impact on many aspects
of this work. Finally, we would like to thank Tim Jones for carefully 
reading this manuscript. E.S and T.P. are grateful to the DESY theory 
group for their kind hospitality.

\clearpage
\providecommand{\href}[2]{#2}\begingroup\raggedright\endgroup

\end{document}